\documentclass[10pt,journal,compsoc]{IEEEtran}
\usepackage{balance}
\usepackage{amsmath,amsfonts}
\usepackage{algorithmic}
\usepackage{algorithm}
\usepackage{array}
\usepackage[caption=false]{subfig}
\usepackage{textcomp}
\usepackage[most]{tcolorbox}
\usepackage{stfloats}
\usepackage{url}
\usepackage{verbatim}
\usepackage{graphicx}
\usepackage{cite}
\usepackage{multirow}
\usepackage{makecell}
\usepackage{balance}
\usepackage{amssymb}
\usepackage{pifont}
\usepackage{xcolor}
\begin{document}
\newcommand{\jl}[1]{{\color{teal}JL: #1}}
\newcommand{\hj}[1]{{\color{green!90}HJ: #1}}
\newcommand{\yb}[1]{{\color{blue!90}YB: #1}}
\newcommand{\rw}[1]{{\color{blue!90}RW: #1}}
\newcommand{\re}[1]{{\color{black}#1}}
\newcommand{\rem}[1]{{\color{black}#1}}

\newcommand{\dataset}{\textsc{\textit{\textbf{KFC}}}}

\title{Evaluating SZZ Implementations:\\ An Empirical Study on the Linux Kernel}

\author{Yunbo Lyu,
        Hong Jin Kang, 
        Ratnadira Widyasari,
        Julia Lawall,
        David Lo~\IEEEmembership{Fellow,~IEEE}
        
\thanks{Y. Lyu, R. Widyasari, D. Lo are with the School of Computing and Information Systems, Singapore Management University. E-mail: \{yunbolyu, davidlo\}@smu.edu.sg, ratnadiraw.2020@phdcs.smu.edu.sg}

\thanks{H.J. Kang is with University of California, Los Angeles. E-mail: hjkang@g.ucla.edu}

\thanks{J. Lawall is with Inria-Paris. E-mail: julia.lawall@inria.fr}
\thanks{Manuscript received April 19, 2021; revised August 16, 2021.}}

\markboth{JOURNAL OF IEEE TRANSACTIONS ON SOFTWARE ENGINEERING, ~Vol.~14, No.~8, August~2023}%
{Shell \MakeLowercase{\textit{et al.}}: A Sample Article Using IEEEtran.cls for IEEE Journals}


\IEEEtitleabstractindextext{
\begin{abstract}
\label{abstract} 
The SZZ algorithm is used to connect bug-fixing commits to the earlier commits that introduced bugs.
This algorithm has many applications and many variants have been devised.
However, there are some types of commits that cannot be traced by the SZZ algorithm, referred to as ``ghost commits".
The evaluation of how these ghost commits impact the SZZ \re{implementations} remains limited.
Moreover, these \re{implementations} have been evaluated on datasets created by software engineering researchers from information in bug trackers and version controlled histories. 

Since Oct 2013, the Linux kernel developers have started labelling bug-fixing patches with the commit identifiers of the corresponding bug-inducing commit(s) as a standard practice. 
As of v6.1-rc5, 76,046 pairs of bug-fixing patches and bug-inducing commits are available. 
This provides a unique opportunity to evaluate the SZZ algorithm on a large dataset that has been created and reviewed by project developers, entirely independently of the biases of software engineering researchers.  

In this paper, we apply six SZZ \re{implementations} 
to 76,046 pairs of bug-fixing patches and bug-introducing commits from the Linux kernel. 
Our findings reveal that SZZ algorithms experience a more significant decline in recall on our dataset ($\downarrow 13.8\%$) as compared to prior findings reported by Rosa et al., and the disparities between the individual SZZ algorithms diminish.
Moreover, we find that 17.47\% of bug-fixing commits are ghost commits.
Finally, we propose Tracing-Commit SZZ (TC-SZZ), that traces all commits in the change history of lines modified or deleted in bug-fixing commits. 
Applying TC-SZZ to all failure cases, excluding ghost commits, we found that TC-SZZ could identify 17.7\% of them.
Our further analysis based on {\em git log} found that 34.6\% of bug-inducing commits were in the function history, 27.5\% in the file history (but not in the function history), and 37.9\% not in the file history.
We further evaluated the effectiveness of ChatGPT in boosting the SZZ algorithm's ability to identify bug-inducing commits in the function history, in the file history and not in the file history.

\end{abstract}

}

\maketitle

\section{Introduction}
\label{sec:introduction}

\IEEEPARstart{I}{n} MSR 2005, {\'S}liwerski, Zimmermann, and Zeller introduced an
approach to identifying bug-inducing commits in a software code base~\cite{sliwerski2005changes}.
Their approach, subsequently referred to as the {\em SZZ algorithm}~\cite{kim2006automatic},
first identifies bug-fixing commits by correlating commits to reports in a bug tracker. 
It then uses the history
mechanism of the version control-system to map each line of code removed by
the bug fix to the commit that most recently added or modified that line.
Finally, the resulting set of commits is {\em filtered} to remove those that postdate the bug report, unless there is evidence that the commit represents a partial fix or the commit is a bug-introducing commit for some other bug-fixing commit.  
The SZZ algorithm has been used in defect prediction techniques~\cite{hata2012bug, tan2015online, pascarella2019fine,
yan2020just, fan2019impact}, and to analyze how bugs are born~\cite{bavota2015four, tufano2017empirical, aman2019empirical, chen2019extracting}.

Despite the success and influence of the SZZ algorithm, over the years, researchers have identified a number of weaknesses in its strategy for identifying bug-inducing commits. 
For example, the SZZ algorithm may incorrectly flag non-semantic source code ({\em e.g.}, formatting changes)\re{~\cite{kim2006automatic}}, 
meta-changes ({\em e.g.}, branch changes)\re{~\cite{da2016framework}}, and refactoring changes\re{~\cite{neto2018impact, neto2019revisiting}} as bug-inducing.
Several algorithm refinements have been proposed, such as excluding formatting and cosmetic changes~\cite{kim2006automatic}, excluding changes that represent version-control system metadata ({\em e.g.}, merge commits) \cite{da2016framework}, excluding changes that perform refactorings~\cite{neto2018impact, neto2019revisiting}, excluding non-executable changes
\cite{williams2008szz}, and combining the syntactic approach proposed by SZZ with analysis of changes in dependencies, including potentially interprocedural analysis~\cite{davies2014comparing}.  

Recently, attention has turned to the notion of a {\em ghost commit}~\cite{rezk2021ghost}. 
As the SZZ algorithm relies on connecting the removed lines in the bug-fixing commit to the added lines in the bug-inducing commit, it is intrinsically unable to provide results for bug-fixing commits that have no removed lines (referred to as {\em Remove Mapping Ghost}), and to detect bug-inducing commits that have no added lines (referred to as {\em \re{Add Mapping Ghost}}).  
The SZZ algorithm furthermore cannot detect bug-inducing commits that are discarded due to the various filtering strategies (referred to as {\em Filtering Ghost}), such as commits dated after the bug report, \re{commits that perform refactoring, or commits that address multiple purposes (i.e., tangled commits~\cite{herzig2013impact})}.
Some further variants have been proposed to mitigate these issues.
\re{Specifically, to address failures from Remove Mapping Ghost, A-SZZ~\cite{sahal2018identifying} and Rezk et al.'s~\cite{da2016framework} method were proposed.
To address failures from Filtering Ghosts, Da Costa et al.~\cite{da2016framework} propose a framework to filter out suspicious commits.}


\textbf{Motivation}: 
While there has been extensive research over two decades aiming to improve the precision of the SZZ algorithm~\cite{kim2006automatic, williams2008szz, davies2014comparing, da2016framework, neto2018impact, neto2019revisiting}, discussions on enhancing recall are relatively limited. 
This analysis gap is critical because overlooking certain buggy patterns can \rem{impact downstream tasks dependent on this algorithm, such as defect prediction~\cite{kim2008classifying, toth2016public, pascarella2019fine, wen2016locus, yan2020just, kim2007predicting, rahman2011bugcache, chen2019extracting} and the analysis of factors related to software quality~\cite{chen2019extracting, aman2019empirical, eyolfson2014correlations, bernardi2018relation, izquierdo2012more, tufano2017empirical}.}
The closest work is that of Sahal et al.~\cite{sahal2018identifying} and Rezk et al.~\cite{rezk2021ghost}, who \rem{use a syntax-based method~\cite{sahal2018identifying} and context-aware data flow analysis~\cite{rezk2021ghost} to address Remove Mapping Ghosts.}
However, their work only partially addresses one possible scenario where SZZ fails to identify bug-inducing commits. 
Moreover, there is a lack of comprehensive examination of the full range of scenarios where the SZZ algorithm falls short\rem{, such as cases where files are not shared between the bug-inducing and bug-fixing commits.}
Therefore, our work aims to present a whole picture of the recall challenges faced by all current SZZ algorithms.


The wide use of the SZZ algorithm and the introduction of its many variants raises the need to evaluate the algorithms fairly.
The original SZZ lacked an empirical evaluation~\cite{sliwerski2005changes}, while variants of SZZ were manually assessed by the authors themselves, rather than by domain experts. 
As noted by da Costa et al.~\cite{da2016framework} and Rosa et al.~\cite{rosa2021evaluating}, due to a lack of developer-labeled data, evaluations of SZZ variants mostly rely on researchers' manual analysis of small datasets~\cite{sliwerski2005changes, kim2006automatic, williams2008szz, davies2014comparing, neto2019revisiting}. 
As the researchers are not developers of the projects being analyzed, their judgement on bug-introducing commits may not always be correct. 
The cost of manual analysis is also overwhelming. 
To avoid the need for manual analysis, while at the same time benefiting from the expertise of project developers, Rosa et al.~\cite{rosa2021evaluating} proposed an automation-based
methodology to build a high-quality dataset from commits where the log message indicates both that the commit fixes a bug and the identity of the bug-inducing commit. 
Still, the resulting dataset only takes a few commits from each of many projects, which might not provide a realistic view.
For instance, in Rosa et al.'s dataset, there exist just three bug-fixing commits from FFmpeg despite FFmpeg having over a hundred thousand commits.


There does exist, however, a large, developer-created dataset linking bug-fixing commits to the corresponding bug-inducing commits, as is needed to evaluate SZZ algorithms.  
Since October 2013, Linux kernel developers have been annotating bug-fixing patches with the keyword ``Fixes:''
followed by the commit id of the bug-inducing commit and the subject-line of the corresponding log message.  
Starting with fewer than 200 such annotated commits in 2013, the practice has steadily grown over the years, until reaching a steady state of 11-13K such commits per year, since 2019.
Kernel developers include {\em Fixes:} annotations in their submitted patches, and the annotations are subject to the kernel code-review process. 
This dataset introduces the possibility of evaluating the SZZ algorithms at a large scale, and accurately measuring the impact of ghost commits.




\textbf{Main Findings}: In this paper, we collect a dataset of commits from the Linux kernel, named the Linux \textbf{K}ernel \textbf{F}ixing \textbf{C}ommits (\dataset{}) dataset. 
The \dataset{} dataset has more than 76K ``Fixes:'' commits over the last nearly 10 years of Linux kernel development to evaluate the SZZ algorithm and five of its variants.  
We find that all of the considered algorithms have recall and precision between 0.40 and 0.60, with some algorithms having higher recall, while others have higher precision, leading to an F1 score in all cases of around 0.50.  
The average recall for all SZZ algorithms evaluated in our dataset is 13.8\% lower than that observed by Rosa et al.\ on their dataset, and the differences between individual SZZ algorithms are less pronounced.

Using the \re{\dataset{}} dataset, we can additionally assess previous work on the frequency of ghost commits.
Rezk {\em et al.}~\cite{rezk2021ghost} have done a large scale study of the prevalence of ghost commits, focusing on 14 Apache projects.
They found a median rate of 7.64\% of occurrences of \re{Remove Mapping Ghost} (bug-fixing commits that \re{have no removed lines}, and thus have no changed lines to map back to bug-inducing commits) over the considered projects.
In the Linux Kernel dataset, we find a much higher percentage, 17.65\%.
We also find a higher percentage than Rezk {\em et al.} of occurrences of Add Mapping Ghost (commits over the whole Linux commit history that \re{have no added lines}), with a rate of 5.3\% for the Linux kernel as compared to a median of 2.68\% for the applications in Rezk {\em et al.}'s dataset.  
Indeed, Rezk {\em et al.} lack a ground truth for the set of bug-inducing commits, and thus compute the rate of all commits that contain no added code as compared to the complete set of commits.
In the case of the Linux kernel, we can also assess the rate of \re{Add Mapping Ghost} commits among the bug-inducing commits labelled by the developer.  
We find that \re{Add Mapping Ghost} bug-inducing commits account for only 0.7\% of all bug-inducing commits (585 out of 79649).  
Thus, the results for the Linux kernel show that the findings of Rezk {\em et al.} across the entire code history do not reflect the actual rate at which commits that \re{have no added lines} are bug-inducing commits. 

We furthermore find that ghost commits are not the only reason for failures (lower recall) of the SZZ algorithms.
Indeed, we find that the bug-inducing commit is often not the most recent previous commit on any of the lines changed by the bug-fixing commit. 
And that over 13\% of bug-inducing commits share no files in common with the subsequent bug-fixing commits.
The files modified in the cross-file bug-inducing commits are present in the version control history but are different from the files in the bug-fixing commits. 
An example of a cross-file bug-fixing commit is commit 79b591c, which does not share any file with its bug-inducing commit. 
The bug-inducing commit modified a constant shared between multiple files, requiring a necessary unit conversion to be applied in each file that uses the constant.  
The bug was introduced as the bug-inducing commit neglected to update a relevant file.
The correct unit conversion in this file was later introduced in the bug-fixing commit.
The SZZ algorithms are founded on the assumption that ``a bug-inducing change adds lines that are later removed by a fix." 
In the context of the git version control system, this means that SZZ algorithms use the {\em git blame} command just once for each removed line to identify the most recent change. 
To increase the recall, admittedly at the expense of precision, we examine whether bug-inducing commits might not always be in the most recent change, but are rather concealed deeper in the change history. 

\begin{figure}[t]
    \centering
    \includegraphics[width=0.48\textwidth]{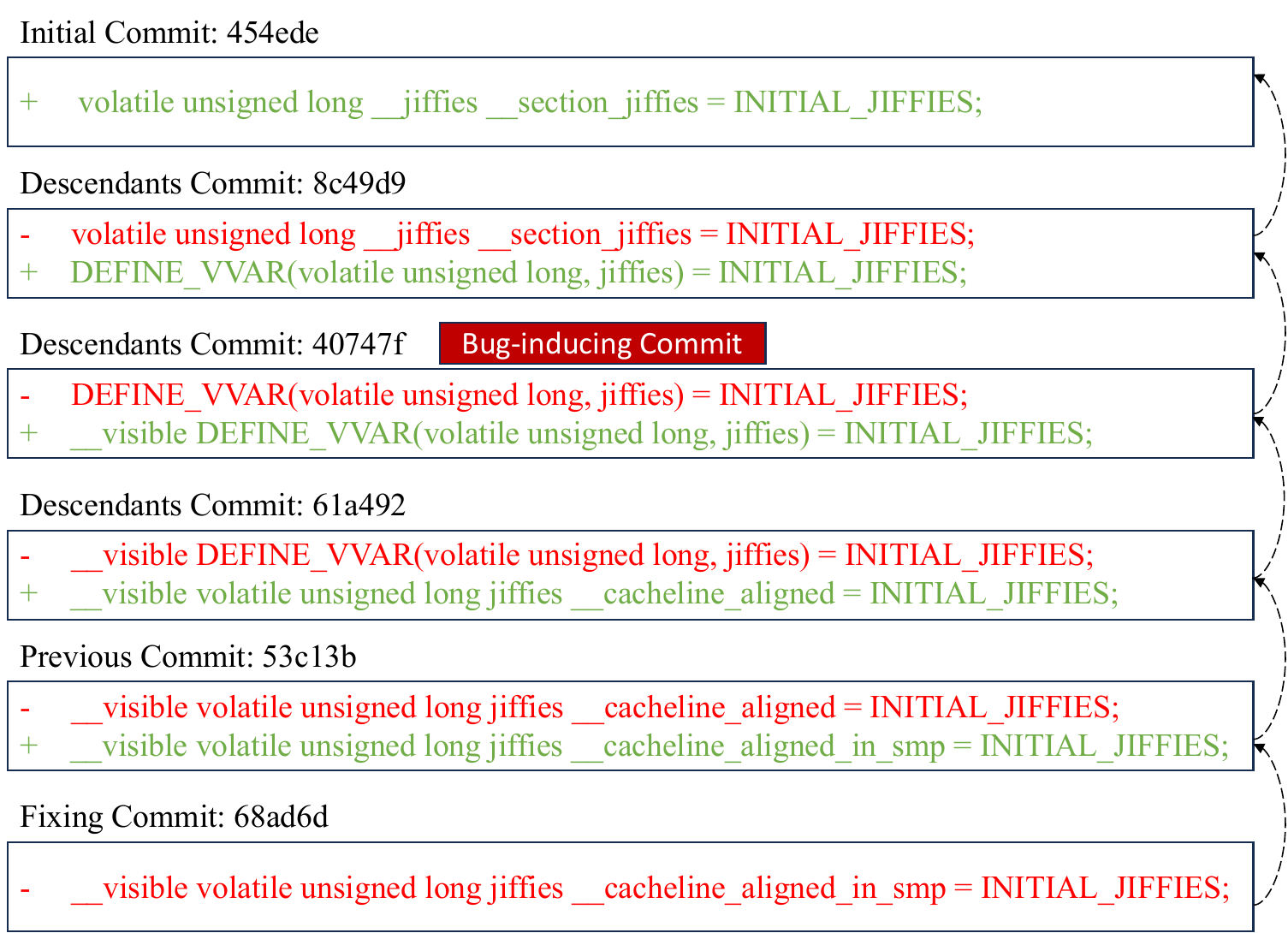}
    \caption{The motivating example of Tracing-Commit SZZ}
    \label{fig:TC-SZZ}
\end{figure}

To facilitate this exploration, we propose the Tracing-Commit SZZ (TC-SZZ) algorithm, which iteratively uses {\em git blame} until the initial commit is located, and then considers all commits in the change history as potential bug-inducing commits.
Unlike all previous SZZ variants which identify only a single bug-inducing commit for a deleted line. 
This favors recall, but may increase the number of false positives as it returns commits in the line's change history that are not relevant to the bug.
\rem{For example, as shown in Fig~\ref{fig:TC-SZZ}, TC-SZZ returns all five commits in a line's change history as potentially bug-inducing commits, successfully identifying commit 40747f as the bug-inducing commit.
In contrast, B-SZZ only considers the previous commit.}
Applying TC-SZZ to the 17,552 cases that are not detected by the SZZ algorithms but do not involve ghost commits revealed that 3,115 out of the 17,552 failure cases (17.7\%) are located in the change history and not the most recent change. 
Algorithms that skip over refactoring changes have the potential to identify such bug introducing commits more accurately, but refactoring detection is language specific, and such approaches have only been developed for Java~\cite{silva2017refdiff, tsantalis2018accurate}, while the Linux kernel is entirely C code.

For the remaining 14,437 bug-fixing commits, 34.6\% of the failure cases are found in the function history but \re{not in the history of the lines that are changed by the bug-fixing commit}. 
\re{This means that the bug-inducing commit can be identified by applying {\em git log -L} on the changed functions (referred to as {\em Function-Log Detection}). 
For detailed examples, refer to Section~\ref{subsec:other_pattern}.
}
Additionally, 27.5\% of failure are found in the file history but not in the history of the changed functions in that file. 
Here, the bug-inducing commit can be found by applying {\em git log} to the changed files (referred to as {\em Within-File Detection}).
Finally, 37.9\% of failure are in the history of a file other than the changed files (referred to as {\em Cross-File Detection}).



\textbf{Experiments with ChatGPT}: For the Function-Log Detection, Within-File Detection, and Cross-File Detection cases, where the bug-inducing commit is not in the history of the lines changed by the bug-fixing commit, we experiment with ChatGPT.
As ChatGPT has exhibited potential in addressing Software Engineering tasks~\cite{zhong2023can, white2023chatgpt}, we consider whether ChatGPT, specifically the GPT-4 version, can help SZZ algorithms identify bug-inducing commits for these failure cases.
Specifically, we use the commit log {\em subject} and {\em body} of the bug-fixing commit, as well as \re{all of the changed} function's code, as input for ChatGPT.
For Within-File Detection and Cross-File Detection, we leverage ChatGPT to help locate the bug-inducing function or file based on the bug-fixing commit, since the function or file responsible for the bug is not known from the bug-fixing commits.
Our aim is for ChatGPT to pinpoint the specific line of code in the function that may have induced the bug and thus caused the bug-fixing commit.
Given this line of code, we can then apply an iterated {\em git blame} to it to identify the bug-inducing commits.
On \re{375} samples, for identifying bug-inducing commits Function-Blame Detection {\color{black}(a subcategory of Function-Log Detection, using {\em git blame} once for identifying the bug-inducing commit)}, ChatGPT showed increased efficacy in recall (with a precision of 0.28, recall of 0.62, and F1 of 0.39). 
However, when it came to dealing with failure cases both in and out of the file history, ChatGPT's performance was notably less successful.
This is partly due to the challenge \rem{of} a general-purpose Large Language Model in inferring function names or files \rem{not explicitly referenced in the bug-fixing commit.}




\textbf{Overview}: In summary, our paper makes the following contributions:
\begin{itemize}
\item \textbf{Dataset Creation.} We identify the Linux kernel as a data source and collect a developer annotated and validated
dataset of over 76K bug-fixing commits, which we have made publicly available.\footnote{https://doi.org/10.6084/m9.figshare.23889792.v2}
\item \textbf{Evaluation.} We conduct a comprehensive evaluation of six SZZ algorithms, revealing hidden dangers in real-world usage scenarios and analyzing the impact of ghost commits on the Linux kernel dataset.
\item \textbf{Further Analysis.} 
We show that by utilizing TC-SZZ, we can address 17.7\%  of the failure cases.
We classified the remaining failure situations as {\em Function-Log Detection}, {\em Within-File Detection}, and {\em Cross-File Detection}. 
\item \textbf{ChatGPT Efficacy Assessment.} 
We investigate the efficacy of ChatGPT in identifying bug-inducing commits in the \re{three} failure situations.
\end{itemize}

The rest of this paper is organized as follows. 
In Section~\ref{sec:background}, we discuss the background and related work of our study. 
In Section~\ref{sec:building}, we discuss how we build the dataset of bug-inducing commits. Section~\ref{sec:design} presents the design of our study. 
Section~\ref{sec:tcszz} introduces TC-SZZ.
In Section~\ref{sec:results}, we show the evaluation results and answer our research questions.
In Section~\ref{sec:discuss}, \re{we explore the potential of ChatGPT to identify bug-inducing commits.}
Section~\ref{sec:reason} discusses the reason of SZZ failures.
We discuss the threats to validity in Section~\ref{sec:threats}. 
Finally, we draw the conclusions of our work and present future directions in Section~\ref{sec:conclusion}.

\section{Background and Related Work}
\label{sec:background}

In this section, we first provide an overview of the original SZZ algorithm, outlining its background and highlighting some limitations related to ghost commits. 
Next, we study the variants of the SZZ algorithm that have been proposed to improve its precision and recall. 
Lastly, we discuss the application and significance of the SZZ algorithm and its variants 
within the Software Engineering research community.

We focus on the eight SZZ algorithms that are considered by Rosa et al.~\cite{rosa2021evaluating}: B-SZZ~\cite{sliwerski2005changes}, AG-SZZ~\cite{kim2006automatic}, DJ-SZZ~\cite{williams2008szz}, L-SZZ~\cite{davies2014comparing},
R-SZZ~\cite{davies2014comparing}, MA-SZZ~\cite{da2016framework}, RA-SZZ~\cite{neto2018impact}, RA-SZZ*~\cite{neto2019revisiting}, as well as some implementation variants: PyDriller (SZZ@PYD)~\cite{spadini2018pydriller}, SZZ Unleashed~\cite{borg2019szz}, and OpenSZZ~\cite{lenarduzzi2020openszz}.

\subsection{The original SZZ algorithm \& Ghost Commits}
\label{subsec:bszz}

SZZ, the name given to the algorithm initially introduced by {\'S}liwerski, Zimmermann, and Zeller~\cite{sliwerski2005changes}, was developed to identify changes that induce bugs. 
In our forthcoming discussion, we will follow Rosa et al.~\cite{rosa2021evaluating} and refer to this original SZZ algorithm as B-SZZ.
B-SZZ follows a core structure that is shared with its subsequent variants:
(1) identifying bug-fixing commits, followed by a filtering process utilizing \rem{two confidence levels}: {\em syntactic} and {\em semantic};
(2) mapping bug-fixing commits to their corresponding potential bug-inducing commits; 
and (3) eliminating potential bug-inducing commits that are unlikely to be the cause of bugs.
 
\subsubsection{Identifying Bug-fixing Commits}
\label{subsubsec:bszz_bic}

In the initial stage, B-SZZ endeavors to establish a relationship between a bug-fixing commit 
in the version control system (VCS) and the corresponding bug reports in the issue-tracking system (ITS). 
To accomplish this, B-SZZ employs two independent confidence levels: a \textit{syntactic} \rem{level}, which infers links between a \re{VCS} log and a bug report, and a \textit{semantic} \rem{level}, which validates these links through bug report data. 
For most software systems, there is no mandatory or default link between bug reports in the ITS and the VCS. Thus, developers may forget to add such links.
Rezk et al.~\cite{rezk2021ghost} define the missing bug-fixing commits that SZZ implementations fail to detect, where links were omitted, as \textbf{Ghost Commit 0 (GC 0)}. 
Herbold et al.~\cite{herbold2022problems} also find that half of the bug-fixing commits identified by SZZ are incorrect due to the mislabeled issues in the ITS (where developers incorrectly label issues as bugs, while some are merely improvements, etc.)~\cite{herzig2013s}.
Herbold et al. also find that SZZ misses about one-fifth of the bug-fixing commits.

This paper will not study the limitations of bug-fixing commit identification, as: 
(1) previous work~\cite{bird2009fair, nguyen2010case, wu2011relink} has already provided a comprehensive analysis and proposed tools to recover GC 0's missing links, 
and (2) the Linux kernel repository requires developers to use the ``Fixes:" tag along with the first 12 characters of the SHA-1 ID when a patch addresses a bug in a specific commit. 
Therefore, we can collect a ``developer-informed" Linux kernel dataset using the VCS without reliance on the ITS.


\subsubsection{Mapping}
\label{subsubsec:bszz_mapping}

Once the bug-fixing commits in the VCS have been identified, B-SZZ proceeds to the second step, where it maps the bug-fixing commits to potential bug-inducing commits, i.e., the commits that may have caused the bug. 
To do this, B-SZZ first employs CVS's {\em diff} command.\footnote{When B-SZZ was proposed, the Concurrent Versions System (CVS) was widely used for version control.} 
This command is analogous to {\em git show},
as it identifies the lines that have been deleted or modified by a commit.
Then, B-SZZ invokes the \textit{annotate} command (another CVS command, similar to {\em git blame},
but that does not identify the corresponding lines in earlier revisions) 
for each line to determine the most recent commit that deleted or modified the line prior to the fix. 
When there are multiple removed or modified lines, multiple commits can be mapped as potential bug-inducing commits.
This paper focuses on issues in the mapping step.

\begin{figure}[ht]
    \centering
    \includegraphics[width=0.45\textwidth]{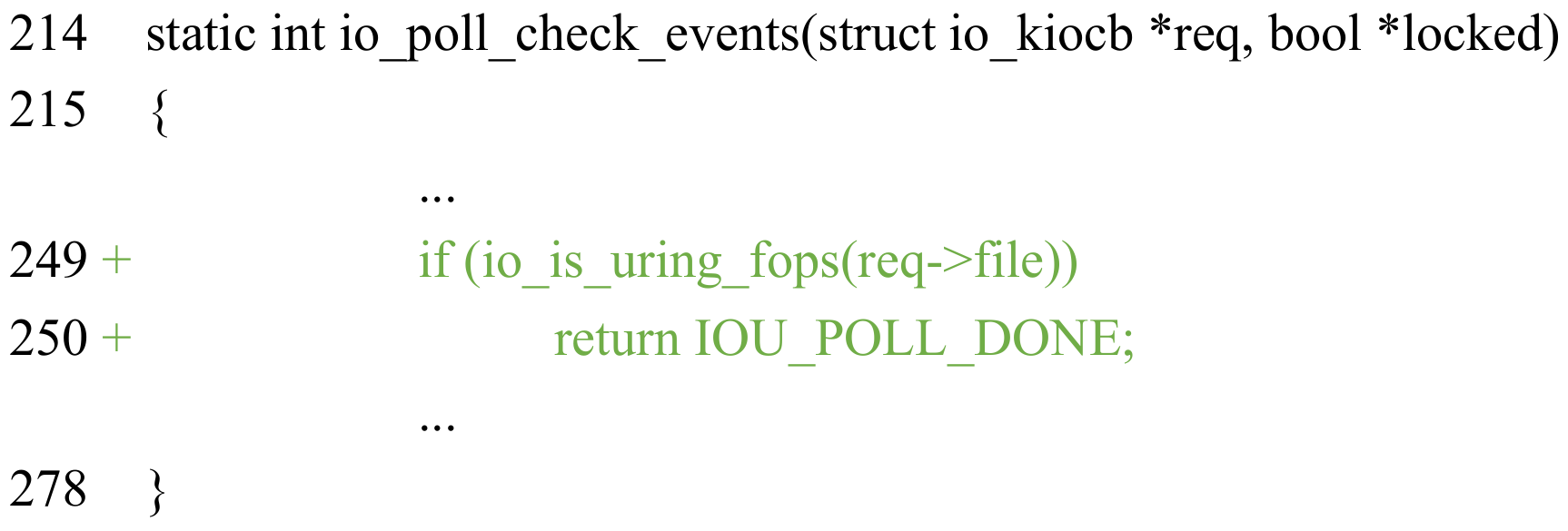}
    \caption{A Remove Mapping Ghost (bug-fixing commit 7fdbc5f).}
    \label{fig:add_mapping_ghost}
\end{figure}

Tools such as CVS \textit{annotate} and \textit{git blame}
cannot map bug-fixing commits that do not modify or remove lines to bug-inducing commits.  
This situation is defined as \textbf{Mapping Ghost type 1 (MG 1)} by Rezk et al.
(To give a more meaningful naming for MG 1, we name it \textbf{Remove Mapping Ghost} in this paper.)
For example, in Fig~\ref{fig:add_mapping_ghost}, bug-fixing commit 7fdbc5f\footnote{https://github.com/torvalds/linux/commit/7fdbc5f014c3f71bc44\\673a2d6c5bb2d12d45f25} does not remove any lines and only added two lines in all the changes.

\begin{figure}[ht]
    \centering
    \includegraphics[width=0.48\textwidth]{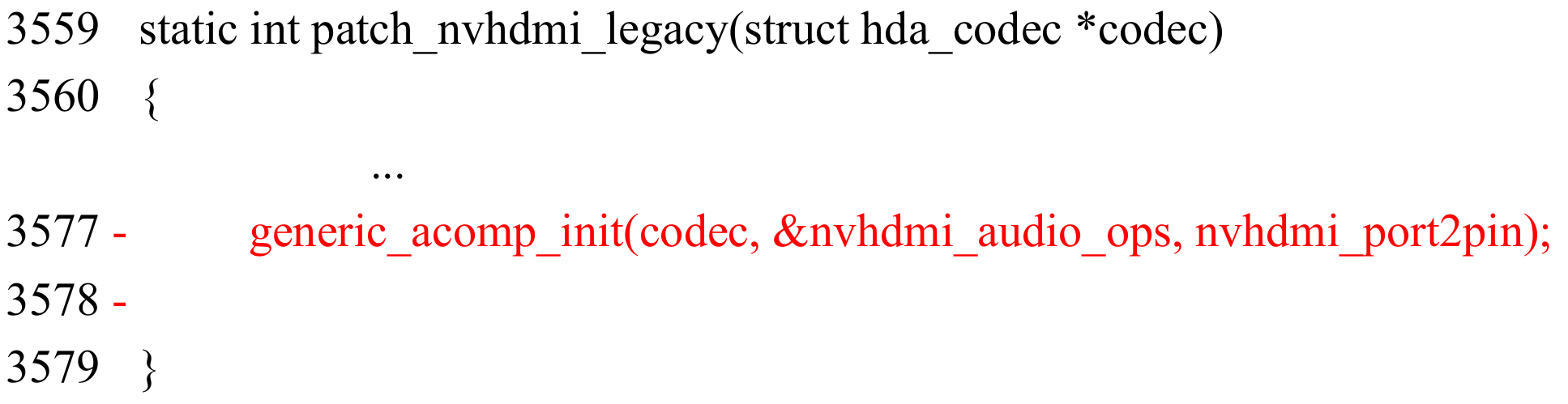}
    \caption{An Add Mapping Ghost (bug-inducing commit 5a858e7).}
    \label{fig:remove_mapping_ghost}
\end{figure}

Similarly, if a bug-inducing commit does not add lines, it cannot be tracked by using {\em blame} on subsequently removed lines, as {\em blame} \re{can only track previously existing lines}. 
This situation is defined as \textbf{Mapping Ghost type 2 (MG 2)} by Rezk et al.
\re{
(To give a more meaningful naming for MG 2, we name it \textbf{Add Mapping Ghost} in this paper.)
For instance, in Fig~\ref{fig:remove_mapping_ghost}, bug-inducing commit 5a858e7\footnote{https://github.com/torvalds/linux/commit/5a858e79c911330678\\b5a9be91a24830e94a0dc9} does not add any lines and only removes two lines in all the changes.
}

\subsubsection{Filtering}
\label{subsubsec:bszz_filtering}

In the final stage, B-SZZ discards the potential bug-inducing commits that cannot have caused the bug, such as those committed to the VCS after the bug was reported. The potential bug-inducing commits that are not filtered out during the filtering phase are reported as bug-inducing commits by B-SZZ.

Subsequent studies have proposed various filtering methods, including Content Filters~\cite{kim2006automatic, mcintosh2018fix} and Suspiciousness Filters~\cite{da2016framework}. If all the potential bug-inducing commits are filtered out for a given bug-fixing commit, Rezk et al. refer to the bug-fixing commit as a \textbf{Filtering Ghost (FG)}. 

We do not further examine the limitations of the filtering step in the SZZ algorithms for two reasons: (1) The filtering step heavily depends on data from the ITS, but we bypass any ITS by using the tags found directly in the Linux kernel commits; (2) The filtering process aims to improve precision by eliminating commits from the potential bug-fixing commits, while our focus is on enhancing the recall of the existing SZZ algorithm, i.e., adding potentially correct bug-inducing commits that were missing.



\subsection{Improvements for B-SZZ}
\label{subsec:szz_precision}

Despite the numerous proposals aimed at enhancing the accuracy of the SZZ algorithm~\cite{kim2006automatic, williams2008szz, davies2014comparing, da2016framework, neto2018impact, neto2019revisiting}, the examination of bug-inducing commits not detected by SZZ is still in its infancy, and only limited efforts have been made to improve the recall~\cite{kim2006automatic, sahal2018identifying, rezk2021ghost}.

Kim et al.~\cite{kim2006automatic} found that B-SZZ may incorrectly label bug-inducing commits due to considering formatting or cosmetic changes. 
To address this, they proposed AG-SZZ, which ignores such changes during the \textit{mapping} process, improving the accuracy of locating bug-introducing commits. 
Moreover, as B-SZZ employs the CVS \textit{annotate} command for mapping lines removed in bug-fixing commits to recent revisions, it cannot pinpoint the corresponding lines in earlier revisions. 
Kim et al. observed that such mappings might lead to inaccurately linked code segments. 
AG-SZZ, therefore, utilizes annotation graphs~\cite{zimmermann2006mining} to map line numbers across revisions, ensuring that both the starting and mapped lines belong to functions with the same name, improving the accuracy of the mapping process.

Williams et al.~\cite{williams2008szz} propose DJ-SZZ to address the limitations of annotation graphs which cannot track individual lines through extensive modifications by using a line-number mapping approach~\cite{williams2008branching}.
Furthermore, they used DiffJ~\cite{pace2007tool} to exclude non-executable changes. 

Davies et al.~\cite{davies2014comparing} propose L-SZZ and R-SZZ, which improve on the AG-SZZ algorithm by selecting only one bug-inducing commit from all the candidate commits. L-SZZ selects the commit that changes the most lines. R-SZZ selects the most recent one. 

Da Costa et al.~\cite{da2016framework} found that AG-SZZ identifies meta-changes (i.e., branch changes, merge changes, and property changes) as bug-inducing commits. 
AG-SZZ includes these commits as the annotation graph is unable to map the code between meta-changes and their preceding changes. 
To address this issue, Da Costa et al. propose MA-SZZ that connects all nodes in the annotation graph to their prior changes. 
This enhancement ensures that MA-SZZ does not identify meta-changes as bug-inducing commits.


Neto et al.~\cite{neto2018impact} found that refactoring operations, which do not modify a program's behavior, impact the SZZ results. 
They extended MA-SZZ as two variants, RA-SZZ~\cite{neto2018impact} and $\text{RA-SZZ}^\text{*}$~\cite{neto2019revisiting}, to reduce the set of false positives reported by MA-SZZ.
RA-SZZ and $\text{RA-SZZ}^\text{*}$ use RefDiff~\cite{silva2017refdiff} and Refactoring Miner~\cite{tsantalis2018accurate}, respectively, to exclude refactoring operations from bug-inducing commit candidates, but both algorithms only work on Java files.

PyDriller (SZZ@PYD)~\cite{spadini2018pydriller}, SZZ Unleashed~\cite{borg2019szz}, and OpenSZZ~\cite{lenarduzzi2020openszz} are three open-source implementations of SZZ algorithms. 
PyDriller is a Python framework that analyzes Git repositories by extracting information about commits, diffs, and other information. 
It also implements B-SZZ and ignores C-style and Python-style comment lines.
SZZ Unleashed uses line-number mapping without DiffJ, which allows it to work on non-Java files. 
OpenSZZ implements B-SZZ by using the \textit{git blame} command. 




To improve the recall of the SZZ algorithm, Sahal et al.~\cite{sahal2018identifying} propose a syntax-based variation, A-SZZ, to address the case of bug-fixing commits that \re{do not have removed lines} (\re{Remove Mapping Ghost}).
This algorithm traces changes for all lines within the code blocks that contain the additions. 
Rezk et al.~\cite{rezk2021ghost} enhance the A-SZZ approach by incorporating context-aware data flow analysis, which includes mitigation strategies for specific programming scenarios such as null checks, creation of new entities, method overrides, logging operations, and class expansions.
These strategies use data flow analysis on identifiers in added code lines, and then use B-SZZ 
to map the lines within these identifiers' data flow paths.

Table~\ref{tab:dataset_size} presents the datasets that have been used for evaluating these algorithms, detailing the dataset type, the number of projects, and the bug-fixing commit count.
In most cases, the ground truth is labeled by researchers without specialized domain knowledge. 
MA-SZZ uses an evaluation framework when there is no ground-truth label available between the bug-fixing commit and the bug-inducing commit. 
Specifically, it uses an automatic metric with three criteria to evaluate SZZ: earliest bug appearance, future impact of a change, and realism of bug introduction~\cite{da2016framework}.
The only dataset labeled by developers contains, on average, only 1.2 commits per project, which may not sufficiently represent the entirety of issues arising in the development process of each project. 
This underlines the significance of collecting a dataset annotated by developers, encapsulating a range of problems that could emerge during a project's lifespan.



\begin{table}[t]
\centering
\caption{Datasets previously used for evaluating the SZZ algorithm. B-SZZ is indicated as //, as the author did not use a dataset for evaluation. ``re." means the dataset is annotated by researchers. ``dev." means the dataset is annotated by developers.} 
\label{tab:dataset_size}
\begin{tabular}{l|lcc}
\hline
     \textbf{Approach / Dataset} & \textbf{\#Dataset type} & \textbf{\#Projects}  
     & \textbf{\#Bug Fixes} \\ \hline
B-SZZ~\cite{sliwerski2005changes}   & //   &  // & //    \\ 
AG-SZZ~\cite{kim2006automatic}  & Manually (re.)   &  2 & 301     \\ 
DJ-SZZ~\cite{williams2008szz} & Manually (re.)   &  1 & 25      \\ 
L-SZZ \& R-SZZ~\cite{davies2014comparing} & Manually (re.)   &  3 & 174      \\ 
MA-SZZ~\cite{da2016framework} & Auto metrics   &  10 & 2,637     \\
RA-SZZ~\cite{neto2018impact}  & Manually (re.)   &  10 & 365\\
$\text{RA-SZZ}^{\ast}$ ~\cite{neto2019revisiting} &  Manually  (re.)   &  10 & 365\\
Rosa's Dataset~\cite{rosa2021evaluating} & Auto (dev.) & 1,625 & 1,930 \\
\hline

\end{tabular}
\end{table}

\subsection{SZZ in SE Research}
\label{subsec:szzinser}

Many prior studies have discussed the significant role that SZZ plays in software engineering research~\cite{rodriguez2018reproducibility, rosa2021evaluating, petrulio2022szz}.

SZZ is used as the first step in many studies, especially in the area of \textbf{defect prediction}. 
Many defect prediction studies use the SZZ algorithm to build datasets to train and evaluate their models~\cite{kim2008classifying, toth2016public, pascarella2019fine, wen2016locus, yan2020just, kim2007predicting, rahman2011bugcache, chen2019extracting}. 
Herbold \textit{et al.}~\cite{herbold2022problems} found that as of August 2019, 8 out of 15 of the public datasets used for defect prediction used B-SZZ to construct their ground truth. However, this approach suggests that the precision of the defect prediction model relies heavily on the quality of the dataset labeled by the SZZ algorithm~\cite{fan2019impact}.
A lot of mislabeled commits in the dataset will seriously affect the model prediction results, which highlights the importance of the accuracy of the SZZ algorithm.

The SZZ algorithm is also used to \textbf{analyze factors related to software quality}~\cite{chen2019extracting, aman2019empirical, eyolfson2014correlations, bernardi2018relation, izquierdo2012more, tufano2017empirical}. 
A study by Eyolfson et al.~\cite{eyolfson2014correlations} discusses the correlation between the time-based factors of a commit (such as the commit's time of day, day of the week, and commit frequency)  and the commit's bugginess. 
Izquierdo-Cort{\'a}zar et al~\cite{izquierdo2012more} and Tufano et al.~\cite{tufano2017empirical} explore the correlation between the developer's experience and the tendency to induce defects. 

SZZ has also been used to identify bug-inducing commits at the pull request level in studies by Petrulio et al.~\cite{petrulio2022szz} and Bludau et al.~\cite{bludau2022pr}. Furthermore, SZZ has been employed to determine the correct version information about vulnerable software within the National Vulnerability Database~\cite{bao2022v}. Given the importance and widespread use of SZZ, we aim to evaluate its performance in our study.

\section{Constructing a dataset of bug-inducing commits informed by developers  }
\label{sec:building}

Evaluating the accuracy of the SZZ algorithm and its variants requires a dataset of bug-fixing commits accompanied in each case by the set of commits that introduced the fixed bug.  {\'S}liwerski, Zimmermann, and
Zeller~\cite{sliwerski2005changes} proposed experiments based on ECLIPSE and MOZILLA in which bug-fixing commits were chosen based on their references to a bug tracker; no evaluation was
made of the accuracy of the corresponding collected bug-inducing commits.
Subsequent SZZ variants were mostly evaluated using small, manually-validated datasets, as shown in Table~\ref{tab:dataset_size}.

\subsection{The dataset of Rosa et al.}

In 2021, Rosa et al.~\cite{rosa2021evaluating} proposed a new methodology to build a
dataset for evaluating SZZ variants based on the observation that the log
messages of some commits both contain a statement that the commit is fixing
a bug {\em and} explicitly identify the commit that introduced the bug that
is being fixed. For example, they have found the message ``THRIFT-4513:
fix bug in comparator introduced by e58f75d'' in the commit a8a97bd in the
project apache/thrift.\footnote{https://github.com/apache/thrift/commit/a8a97bde9eeada5ce05\\71ea1650b18f3ebc50e42}
The presence of such messages raises the possibility of creating a large-scale dataset fully
automatically, relying on the expertise of project developers to map bug-fixing commits to the corresponding set of bug-introducing commits.
Nevertheless, they did not identify a standardized means of providing bug-fixing and bug-introducing commit information neither across projects nor within projects.  
Thus, they proposed to mine such information in commit logs using natural language processing (NLP) techniques, followed by manual verification.  Using their approach, they created a dataset of 1,930 bug-fixing commits from 1,625 software projects written in a variety of programming languages.  
Using the resulting dataset, they compare the precision, recall, and F-measure of \rem{nine} SZZ variants, and study the impact of dropping reported bug-inducing commits that postdate the associated entry in the bug tracker.

While the NLP-based approach of Rosa et al.~\cite{rosa2021evaluating} made it possible to obtain a
large dataset, exploiting developer knowledge, researchers have identified some weaknesses in their results.  
The dataset remains small, containing
1,930 bug-fixing commits from 19,603,736 commits collected on GitHub, amounting to
0.001\% of the total commits collected.  While not all commits on GitHub are fixes or introduce bugs, previous studies~\cite{erlikh2000leveraging, minelli2015know} have highlighted that 55\% to 95\% of the total cost of software development is devoted to maintenance.
Therefore, it is likely that the 1,930 commits collected in the dataset is a vast underestimation of the actual number of bug fixes.
And as there are on average only 1.2 commits per project, the collected
commits may not be representative of the full scope of issues that arise in
each project's development.

\subsection{A dataset extracted from the development history of the Linux kernel}

The Linux kernel was first released in 1991 and has grown to 23 MLOC as of
the release of Linux v6.0 in October 2022.  The Linux kernel has been
developed using \texttt{git} since April 2005.  Prior to October 2013,
Linux kernel developers sporadically included in their commit log messages
comments beginning with {\bf Fixes:} followed by a reference to Bugzilla or
some other URL.  
Starting with the commit 629c66a22c21\footnote{https://github.com/torvalds/linux/commit/629c66a22c21b692b6\\e58b9c1d8fa56a60ccb52d} from security researcher Kees Cook on October 24, 2013, the Linux
kernel developer community has mostly settled on a strict format for
bug-fix information. This format consists of one or more lines starting with the
keyword \textbf{Fixes:}, each followed by the commit id of the bug-introducing
commit, followed by the first line of the commit log message of the
bug-introducing commit in parentheses.  
This format is \re{encouraged} by the Linux kernel documentation\footnote{https://www.kernel.org/doc/html/v4.9/kernel-documentation.html} and by the Linux
kernel ``linting'' tool \texttt{checkpatch}.\footnote{https://docs.kernel.org/dev-tools/checkpatch.html}

Using \texttt{git}, we collect 78,384 potential bug-fixing commits with \textbf{Fixes:}. 
We further mine all bug-inducing commits in each bug-fixing commit. 
In the first step, we use \textit{[0-9a-f]\{7,40\}} to find the partial commit IDs (7-40 digits) of potential bug-inducing commits. 
We exclude partial commit IDs
with fewer than \rem{seven} characters to prevent ambiguity. 
In the second step, we search for the partial commit IDs in the list of actual complete commit IDs in the repository.
If the partial commit ID can be found anywhere as a substring of a complete commit ID, we return the complete commit ID as the bug-inducing commit. 
To investigate the number of false positives in this step, we compare our results with those obtained using the {\em git rev-parse} command, \rem{which returns the full hash from a partial commit ID.}\footnote{https://git-scm.com/docs/git-rev-parse}
\rem{As a result of this comparative analysis (of 104 cases), we observed two main issues that account for the differences between our method and the {\em git rev-parse} command}:
(1) \textbf{Overlapping Partial Commit IDs}: In certain instances, a partial commit ID is shared by multiple complete commit IDs, leading to false positives. 
\rem{We manually verified such cases and corrected 51.} 
(2) \textbf{Typos in Commit IDs}: 
We found that the remaining 55 cases were the result of typos in commit IDs.
Occasionally, commit IDs in the Linux kernel are incorrect due to typos, such as missing characters, which \rem{prevents} {\em git rev-parse} from identifying the correct commit. 
For example, the partial commit ID af4df655040 should actually be 3af4df655040, making {\em git rev-parse} unable to find 3af4df65504088e9a7d20c0251e1016e521ad4fc. 
Searching for partial commit IDs in the list accurately identified 55 of 56 cases. 
We encountered one instance where the commit did not belong to any branch in the repository, being part of a fork outside the repository. 
In the last step, we collect all the potential bug-fixing commits for which we can find at least one bug-inducing commit into the \textbf{Bug-fixing Commit Dataset}. 
We then collect those commits for which we cannot find any bug-inducing commits into the CSV file included in the replication package.

Our resulting Bug-fixing Commit Dataset contains 
76,046 commits from between Oct 24, 2013 and Nov
20, 2022 that contain one or more properly formatted {\bf Fixes:}
annotations.  
The Linux kernel does not have an associated bug tracker through which we can validate that the annotated commits are fixing bugs.  
Nevertheless, the {\bf Fixes:} annotation has, like the rest of the commit, been subject to the Linux kernel review process.  
Furthermore, we find that over 99\% of the {\bf Fixes:} annotations refer to existing commits.

\re{
We name our newly collected dataset \dataset{} (Linux \textbf{K}ernel Bug-\textbf{F}ixing \textbf{C}ommits).
\dataset{} contains a one-to-many relationship between bug-inducing commits and bug-fixing commits, as a single bug-inducing commit can lead to multiple future bug-fixing commits. 
Conversely, there is also a many-to-one relationship, where a single bug-fixing commit may be the result of multiple bug-inducing commits. 
}

\subsection{Verification of the Dataset}\label{subsec:verify}

To verify the quality of our dataset, we conduct a manual analysis on a subset of the \dataset{} dataset. 
This analysis aims to confirm that the ``Fixes:" field is indeed used in practice to point to the cause of the bug, rather than mistakenly pointing to the initial implementation of related functionalities.

In this study, we followed four steps to validate the subset of the \dataset{} dataset:
\begin{enumerate}
    \renewcommand{\labelenumi}{(\roman{enumi})} 
    \item {\em Sampling}: 
    The complete \dataset{} dataset includes 76,046 bug-fixing commits. 
    We selected a random sample of 383 cases to provide a statistically significant representation, aiming for a 95\% confidence level with a 5\% margin of error, according to standard sample size calculations.\footnote{https://www.surveymonkey.com/mp/sample-size-calculator}
    \item {\em Preparation}: 
    Three annotators initially discussed and refined their approach to ensure consistent labeling, based on an examination of a preliminary set of examples. 
    This led to the creation of a comprehensive set of labeling guidelines.
    \item {\em Labeling}: Following these guidelines, the dataset was then labeled. One annotator was responsible for labeling all 383 cases, while the other two split the complete set, each labeling either 191 or 192 cases.
    \item {\em Discussion}: Upon completion of labeling, the annotators met to discuss and resolve all discrepancies. 
\end{enumerate}

After the {\em preparation} step, six categories were established for the labeling process:
(1) ``Yes": The bug-inducing commit clearly points to the cause of the bug.
(2) ``Affected by other changes": External modifications led to the emergence of the bug.
(3) ``Partial fix": An intermediate commit attempted to fix the bug but did not completely succeed.
(4)  ``Introducing a new function": The commit introduced new functionality that was later involved in a bug, but was not directly the cause.
(5) ``No": The bug-inducing commit does not point to the cause of the bug and does not fit into categories (2)-(4). Annotators are required to mark the reasoning behind this classification.
(6) ``Not sure": The annotator could not determine whether the bug-inducing commit is related to the cause of the bug and is instructed to mark the reasons for this uncertainty.

After the {\em Labeling} and {\em Discussion} steps, we found that 81\% (317 out of 383) of the sample cases fell into the ``Yes" category.
Additionally, classifications for the remaining cases were distributed as follows: 
4\% (15 out of 383) were categorized as``Introducing a new function,"  2\% (9 out of 383) as ``Affected by other changes," and 3\% (11 out of 383) as ``Partial fix."
A further 2\% (9 out of 383) of cases were labeled as ``No," as the annotators could not discern a relationship between the bug-fixing and bug-inducing commits.
Lastly, 7\% (26 out of 383) of the cases were tagged as ``Not sure," due to the annotators' lack of expertise in the Linux kernel, which prevented them from making judgments.
Despite a small percentage of cases not directly representing the cause of the bug, the vast majority of the dataset accurately reflects the intended use of the "Fixes:" field.

\section{Study Design}\label{sec:design}

The purpose of this study is to empirically evaluate the various implementations of the SZZ algorithm on the currently largest dataset for SZZ evaluation and to investigate patterns that cannot be identified by SZZ (i.e., how to improve the recall of the SZZ algorithms), particularly Ghost Commits. To achieve these goals, we formulate three research questions:

\textbf{\underline{RQ1:} How well do the different variants of SZZ perform in identifying bug-inducing changes?}  

In our first research question, we aim to examine the performance of different SZZ variants on our dataset relative to a previous study by Rosa et al.~\cite{rosa2021evaluating}. Our study involves a dataset with more bug-fixing commits than in previous studies. Evaluating SZZ is crucial to ascertain if the prior studies' claims remain valid in an updated and more comprehensive dataset from a \textbf{complete} software project. 

\textbf{\underline{RQ2:} What is the impact of Ghost Commits on the behavior of SZZ in a real software project?}  

In our second research question, we aim to explore the effects of ghost commits on the use of SZZ variants in a real software project, 
We divide RQ2 into two specific questions: 
(1) How frequently do ghost commits (Remove Mapping Ghost as bug-fixing commits, Add Mapping Ghost as bug-inducing commits) 
occur? 
(2) How do ghost commits impact the SZZ algorithms? 

\textbf{\underline{RQ3:} What other situations, besides Ghost Commits, are undetectable by SZZ algorithms?}  
In our final research question, we aim to identify additional situations where the SZZ algorithms fail, apart from the Ghost Commit situation. 
Given that SZZ algorithms are founded on the assumption that ``a bug-inducing change adds lines that are later removed by a fix,'' 
we address two 
specific questions within RQ3: 
(1) Do all failure cases, besides Ghost Commits, adhere to this assumption? 
(2) Beyond the previously studied situation, what other situations prevent SZZ from successfully identifying bug-inducing commits?

\subsection{Algorithm Selection \& Data Collection}
\label{subsec:data_collection}

We conduct experiments on several variants of the SZZ algorithm. 
Specifically, we adapt the main variants from the prior study by Rosa et al.~\cite{rosa2021evaluating} to run on our dataset.
Of these, we do not use SZZ Unleashed since our commits are not associated with an issue tracker; thus there is no issue date. 
We do not use OpenSZZ since it needs a Jira URL as input. 
$\text{RA-SZZ}^{\ast}$ only works on Java as Refactoring Miner \cite{tsantalis2018accurate} can only detect refactoring operations in Java files.
Therefore we do not use $\text{RA-SZZ}^{\ast}$
because the Linux kernel is a C-based project. 
In summary, we use B-SZZ, AG-SZZ, L-SZZ, R-SZZ, MA-SZZ, and PyDriller (SZZ@PYD). 
We used the \re{\dataset{} dataset}, featuring the complete set of 76,046 {\bf Fixes:} annotated bug-fixes.


\subsection{Data Analysis}
\label{subsec:data_analysis}
Given the set of bug-inducing commits predicted by the tested SZZ algorithms and the collected dataset, we evaluate the accuracy of the tested SZZ algorithms using recall and precision, two well-known Information Retrieval metrics~\cite{baeza1999modern}. We also use F1 score~\cite{manning2008introduction}, which calculates the harmonic mean of recall and precision. These metrics have been used in previous studies to evaluate SZZ performance~\cite{rosa2021evaluating, petrulio2022szz}.  
The metrics for a bug-fixing commits are defined as follows, where $\mathit{correct_{c_i}}$ is the ground truth of bug-fixing commit $c_i$, i.e., the bug-inducing commits annotated by the original Linux kernel developer, $\mathit{identified_{c_i}}$ is the bug-inducing commits of bug-fixing commit $c_i$ identified by the SZZ algorithms,
and $N$ is the total number of bug-fixing commits: 


\begin{equation*}
    \begin{array}{c}
    \mathit{recall} = \dfrac{1}{N} \sum\limits_{c_i=1}^{N} \dfrac{|\mathit{correct_{c_i}} \cap \mathit{identified_{c_i}|}}{|\mathit{correct_{c_i}|}}
    \end{array}
\end{equation*}

\begin{equation*}
    \begin{array}{c}
    \mathit{precision} = \dfrac{1}{N} \sum\limits_{c_i=1}^{N} \dfrac{|\mathit{correct_{c_i}} \cap \mathit{identified_{c_i}}|}{|\mathit{identified_{c_i}}|}  
    \end{array}
\end{equation*}

\begin{equation*}
    \begin{array}{c}
    F1 = 2 \cdot \dfrac{recall \times precision}{recall + precision}
    \end{array}
\end{equation*}


We also analyze the complementarity of the SZZ variants following previous work~\cite{rosa2021evaluating}. 
Given the set of evaluated 
SZZ variants $\mathit{SZZ}_{eval} = \{v_1, v_2, \ldots, v_n\}$, we compute their complementarity using the following metrics for each $v_i$~\cite{oliveto2010equivalence}:

\begin{equation*}
    \begin{array}{c}
    \mathit{correct}_{v_i \cap v_j}  = \dfrac{|\mathit{correct}_{v_i} \cap \mathit{correct}_{v_j}|} {|\mathit{correct}_{v_i} \cup \mathit{correct}_{v_j}|}
    \end{array}
\end{equation*}

\begin{equation*}
    \begin{array}{c}
    \mathit{correct}_{v_i \setminus (\mathit{SZZ}_{exp} \setminus v_i)} = 
    \dfrac{|\mathit{correct}_{v_i} \setminus \mathit{correct}_{(\mathit{SZZ}_{exp} \setminus v_i)}|}
    {|\mathit{correct}_{v_i} \cup \mathit{correct}_{(\mathit{SZZ}_{exp} \setminus v_i)}|}
    \end{array}
\end{equation*}

\noindent  $\mathit{correct}_{v_i}$ denotes the set of correct bug-inducing commits predicted by the SZZ algorithm $v_i$. $\mathit{correct}_{(SZZexp \setminus v_i)}$ denotes the correct bug-inducing commits predicted by all evaluated SZZ algorithms except for $v_i$. 
$\mathit{correct}_{v_i \cap v_j}$ represents the overlap between the correct bug-inducing commits identified by the SZZ algorithm $\mathit{v_i}$ and another SZZ algorithm $\mathit{v_j}$.
$\mathit{correct}_{v_i \setminus (SZZ_{exp} \setminus v_i)}$, denotes the correct bug-inducing commits that are only identified by the SZZ algorithm $v_i$.




To analyze the frequency of each \re{Remove Mapping Ghost} and \re{Add Mapping Ghost}, we use the same computation as Rezk et al.~\cite{rezk2021ghost}. 
\re{Remove Mapping Ghost}: the percentage of bug-fixing commits that do not remove lines and only add lines out of all bug-fixing commits. 
\re{Add Mapping Ghost}: the percentage of commits that do not add lines and only remove lines out of {\em all} commits. 
With respect to \re{Add Mapping Ghost}, we also report the percentage of bug-inducing commits that \re{do not add lines} and only remove lines out of the set of bug-inducing commits, as these are the only commits that only remove lines that actually have an impact on the SZZ results.

\section{TC-SZZ}\label{sec:tcszz}

The fundamental premise 
of the SZZ algorithm is that ``a bug-inducing change adds lines that are later removed by a fix." 
In other words, for a bug-fixing commit, the SZZ algorithms use the \textit{git show} command 
to identify deleted or modified lines, and the \textit{git blame} command to identify the most recent change to each such line. 
However, we discovered that some bug-inducing commits are not the most recent changes before the bug fix, but rather appeared earlier in a deleted or modified line's change history. 

For example, Fig~\ref{fig:TC-SZZ} shows a line's change history (i.e., iteratively using {\em git blame} until the initial commit) of the bug-fixing commit 68ad6d.\footnote{https://github.com/torvalds/linux/commit/d8ad6d39c35d2b44b\\3d48b787df7f3359381dcbf}
Following Bao et al.~\cite{bao2022v}, we call the immediately preceding the lines changed in the fixing commit the \textbf{previous commits} (i.e., commit 53c13b in Fig~\ref{fig:TC-SZZ}) and all commits (except the initial one) that previously modified these lines as the \textbf{descendant commits} (i.e., commit 8c49d9, 40747f, 61a492 in Fig~\ref{fig:TC-SZZ}). 
We refer to the commit at which tracing can no longer continue as the \textbf{initial commit} (i.e., commit 454ede in Fig~\ref{fig:TC-SZZ}).
\rem{B-SZZ} identified the {\em previous commit} as the bug-inducing commit, relying on using {\em git blame} once.
\rem{
Other algorithms, such as AG-SZZ and MA-SZZ, which improve on B-SZZ, access the rest of the descendant commits, identifying bug-inducing commits from other commits by considering criteria such as whether the commit is a meta-change. 
However, no existing algorithm comprehensively examines the line change history to identify the bug-inducing commit.
}



Therefore, we propose the Tracing-Commit-SZZ (TC-SZZ) algorithm, which assumes a distinct function
as compared to previous SZZ algorithms. 
Unlike the six SZZ algorithms discussed in this paper, which \rem{normally} identify only the {\em previous commit} as the bug-inducing commit, and V-SZZ~\cite{bao2022v}, which considers the {\em initial commit} based on the observation
that vulnerabilities are often foundational, TC-SZZ is designed to more effectively locate and understand the position of the bug-inducing commit in the change history of lines.
TC-SZZ applies {\em git blame} iteratively to \re{all deleted or modified lines in the bug-fixing commit} until it reaches the {\em initial commit}, and returns all commits (i.e., the descendant commits and the initial commit in Fig \ref{fig:TC-SZZ}) in the change history of these lines.
If {\em git blame} leads to a commit that has multiple removed lines,
TC-SZZ chooses one among them to continue the iteration.
Choosing all lines could lead to too much fanout, giving many false positives.
For this, we have developed an algorithm based on V-SZZ~\cite{bao2022v}.  
Our algorithm takes into account both the string similarity and the relative distance between the line reported by {\em git blame} and the deleted lines in the same commit.  

An alternative to the iterative use of git blame would be to use {\em git log -L}, which tracks changes to specific lines through the history.
TC-SZZ, however, adds an additional dimension by considering string similarity, which can account for moved lines, thus identifying more bug-inducing commits than {\em git log -L}. 
Take for example commit 5b47348.\footnote{https://github.com/torvalds/linux/commit/5b47348fc0b18a78c9\\6f8474cc90b7525ad1bbfe} 
If we apply the {\em git log -L} command on the lines deleted by this bug-fixing commit, we only trace back to commit ed928a3.\footnote{https://github.com/torvalds/linux/commit/ed928a3402d8a24a7\\0c242c63109c069a7b1f3ab}
In this commit, the developer relocated a few functions to a different part of the file. 
The {\em git log -L} command interprets this as the initial commit for these lines. 
However, TC-SZZ goes beyond this by finding lines with similar content and executing {\em git blame} once more, enabling us to identify the bug-inducing commits.

To facilitate future research on understanding the placement of bug-inducing commits in the change history, TC-SZZ offers multiple modes:
(1) \textbf{Chronological Trace Mode}: For every deleted or modified line in a bug-fixing commit, TC-SZZ returns all potential bug-inducing commits in the order of {\em git blame}. 
For instance, Figure \ref{fig:TC-SZZ} illustrates the change history of a bug-fixing commit for a specific deleted or modified line, with TC-SZZ providing the sequence: \re{previous commit, descendants commit 61a492, 40747f, 8c49d9, and initial commit}.
(2) \textbf{Unique Commits Mode}: For the sake of conciseness, based on the results from Chronological Trace Mode, only unique potential bug-inducing commits are returned, eliminating duplicates.
(3) \textbf{Custom Blame Invocation Mode}: The user has the ability to specify the number of times {\em git blame} is invoked (a parameter of 1 is standard for SZZ, a parameter of 2 means using {\em git blame} twice for each modified line, and a parameter of -1 identifies the initial commit).
Our implementation of TC-SZZ is publicly available in the replication package.

\section{Results}\label{sec:results}
In this section, we answer the three Research Questions proposed in Section~\ref{sec:design}.

\subsection{RQ1: SZZ Performance} \label{subsec:szz_performance}

Table \ref{tab:dataset_new} presents the results for the six SZZ algorithms, ordered chronologically, on our newly collected \re{\dataset{} dataset}, compared against the results reported by Rosa et al.~\cite{rosa2021evaluating} for their dataset ($\text{dataset}_{\text{rosa}}$).

\begin{table}[!t]
\centering
\caption{Recall, Precision, and F1 results of all SZZ algorithms.}
\label{tab:dataset_new}
\begin{tabular}{@{}lcccccc@{}}
\hline
\multirowcell{2}{\textbf{Algorithm}} &\multicolumn{3}{c}{$\text{dataset}_{\text{linux}}$}  &\multicolumn{3}{c}{$\text{dataset}_{\text{rosa}}$}\\ & \textbf{Precision} & \textbf{Recall}  & \textbf{F1}  & \textbf{Precision}  & \textbf{Recall} & \textbf{F1} \\ \hline
B-SZZ  & 0.42 & 0.58 & 0.49 & 0.39 & 0.69 & 0.50\\ 
AG-SZZ & 0.42  & 0.56 & 0.48 & 0.45 & 0.60 & 0.52 \\ 
L-SZZ  & 0.57 & 0.43 & 0.49 & 0.52 & 0.45 & 0.48 \\ 
R-SZZ  & 0.59 & 0.45 & 0.51 & 0.66 & 0.57 & 0.61 \\
MA-SZZ & 0.40 & 0.55 & 0.46 & 0.36 & 0.64 & 0.46 \\
SZZ@PYD & 0.43 & 0.55 & 0.48 & 0.39 & 0.67 & 0.49 \\
\hline
\end{tabular}
\end{table}


The results demonstrate that R-SZZ is the best algorithm (with the highest F1) on both datasets, given that F1 combines precision and recall. 
There are several factors that cause R-SZZ to outperform the other SZZ variants on our dataset.
First and foremost, R-SZZ only returns the most recent commit among all candidates. 
At the same time, 96.1\% (73,109) of the Linux bug-fixing commits have only one bug-inducing commit.
Thus, R-SZZ, by returning only one bug-inducing commit, achieves higher precision than SZZ algorithms that return multiple bug-inducing commits.

Four of the six algorithms demonstrate better precision on the \dataset{} dataset than on the baseline ({\em i.e.}, Rosa's result),
with an average improvement of 9.67\% for these four algorithms. 
Only AG-SZZ and R-SZZ have lower precision on the \dataset{} dataset than on the baseline. 
R-SZZ and L-SZZ hold a significant advantage in precision compared to other SZZ variants on the \dataset{} dataset.
AG-SZZ, the third-best performer, performs in line with B-SZZ on the Linux dataset (42\%). SZZ@PYD ranks as the fourth best algorithm in Rosa's dataset, but it is the third best algorithm in our dataset. 
Concurrently, MA-SZZ consistently underperforms. 


\begin{table}[t]
\centering
\caption{Average number of bug-inducing commits identified by various SZZ algorithms per bug-fixing commit.}
\label{tab:potential}
\begin{tabular}{lll}
\hline
\textbf{Algorithm} &  $\text{dataset}_{\text{linux}}$ & $\text{dataset}_{\text{rosa}}$ \\ \hline
B-SZZ & 1.76 & 1.96 \\ 
AG-SZZ & 1.73 & 1.73 \\ 
L-SZZ  & 1 & 1 \\ 
R-SZZ  & 1 & 1 \\
MA-SZZ & 1.81 & 2.01\\
SZZ@PYD & 1.73 & 1.95\\
\hline
\end{tabular}
\end{table}


Table~\ref{tab:potential} presents the average number of bug-inducing commits identified by the different SZZ algorithms for each bug-fixing commit across the two datasets, excluding the Ghost Commits.
For instance, for each bug-fixing commit on Rosa's dataset~\cite{rosa2021evaluating}, B-SZZ identifies an average of 1.96 bug-inducing commits, whereas, on the \dataset{} dataset, B-SZZ only identifies an average of 1.76 bug-inducing commits. 
Thus, the precision of B-SZZ of 0.39 on Rosa's dataset increases to 0.42 on the Linux kernel dataset, an increase of 7.7\%.
The results in Table \ref{tab:dataset_new} and Table \ref{tab:potential} clearly show
that as the number of identified commits decreases, precision increases.



\underline{Recall:} 
As shown in Table~\ref{tab:dataset_new}, all SZZ algorithms exhibit varying degrees of decline in recall as compared to the results on Rosa's Dataset, with an average decrease of 13.34\% across the six algorithms. 
This highlights the challenge of achieving good recall with the SZZ algorithm, an issue that seems more significant on the Linux kernel dataset than in the previous empirical study~\cite{rosa2021evaluating}.
Among all SZZ algorithms, B-SZZ consistently delivers the best performance on both datasets. In contrast, L-SZZ and R-SZZ are the least effective algorithms, as they identify only one potential bug-inducing commit each. MA-SZZ consistently ranks as the third-best algorithm. Notably, MA-SZZ, AG-SZZ, and SZZ@PYD demonstrate remarkably similar performance on the Linux dataset, with a mere 0.1\% difference between them. 
One source of missed bug-inducing commits is ghost commits.
In Section~\ref{subsec:ghost_impact}, we will examine the influence of ghost commits on the recall achieved by the SZZ algorithms. 
Then, in Section~\ref{subsec:other_pattern}, we will investigate the other situations SZZ fails to identify bug-inducing commits.

Our findings confirm that the top-performing algorithms for precision and recall remain R-SZZ ($\sim$59\%) and B-SZZ ($\sim$58\%), respectively, which validates the previous research by Rosa et al. 
However, when applied to a complete repository such as the Linux kernel, the precision of R-SZZ and recall of B-SZZ, two leading SZZ algorithms, decline by 10.6\% and 15.9\%, respectively. 
While R-SZZ can still be useful for cases where precision is crucial and B-SZZ when recall is crucial, it is important to note that both algorithms still have limitations in accurately and comprehensively identifying bug-inducing commits in software repositories. 

\begin{table}[t]
\centering
\caption{Bug-inducing commits correctly identified exclusively by each SZZ algorithm.}
\label{tab:unique}
\begin{tabular}{l|rl|rl}
\hline
\textbf{Algorithm} &  $\text{dataset}_{\text{linux}}$ & percentage & $\text{dataset}_{\text{rosa}}$ & percentage\\ \hline
B-SZZ  & 1024/47352 & 2.163\% & 21/784 & 2.679\%\\ 
AG-SZZ & 2/47352 & 0.004\%    & 0/784  & 0\%\\ 
L-SZZ  & 0/47352 & 0\%        & 0/784  & 0\%\\ 
R-SZZ  & 0/47352 & 0\%        & 0/784  & 0\%\\
MA-SZZ & 160/47352 & 0.338\%  & 1/784  & 0.128\%\\
SZZ@PYD & 66/47352 &  0.139\% & 0/784  & 0\%\\
\hline
\end{tabular}
\end{table}

Table~\ref{tab:unique} presents the $\mathit{correct}_{v_i \setminus (SZZ_{exp} \setminus v_i)}$ metric calculated for each SZZ variant $v_i$. 
This metric represents the correct bug-inducing commits identified exclusively by $v_i$ and not by the other SZZ algorithms. 
\re{
For instance, consider the \dataset{} dataset, where all SZZ algorithms combined correctly identify a total of 47,352 bug-inducing commits. 
Of these, B-SZZ uniquely and correctly identifies 1,024 bug-inducing commits not identified by other SZZ algorithms. 
Similarly, in the $\text{dataset}_{\text{rosa}}$, the total number of correctly identified bug-inducing commits by all SZZ algorithms is 784, with B-SZZ alone correctly identifying 21 bug-inducing commits that the other algorithms did not.
}
In $\text{dataset}_{\text{linux}}$, four techniques (i.e., B-SZZ, AG-SZZ, MA-SZZ, SZZ@PYD) identify bug-inducing commits that remain undiscovered by other techniques, while in $\text{dataset}_{\text{rosa}}$, only two techniques (B-SZZ, MA-SZZ) exhibit this ability.
Given that B-SZZ does not filter any potential bug-inducing commits and shows the highest recall in the results displayed in Table~\ref{tab:dataset_new}, it has the most uniquely identified bug-inducing commits across both datasets. Interestingly, despite having the third-lowest recall, MA-SZZ can uniquely identify 160 bug-inducing commits.

\begin{figure*}[htbp]
    \centering
    \subfloat[\dataset{} dataset\label{fig:heatmap}]{
        \includegraphics[width=0.45\textwidth]{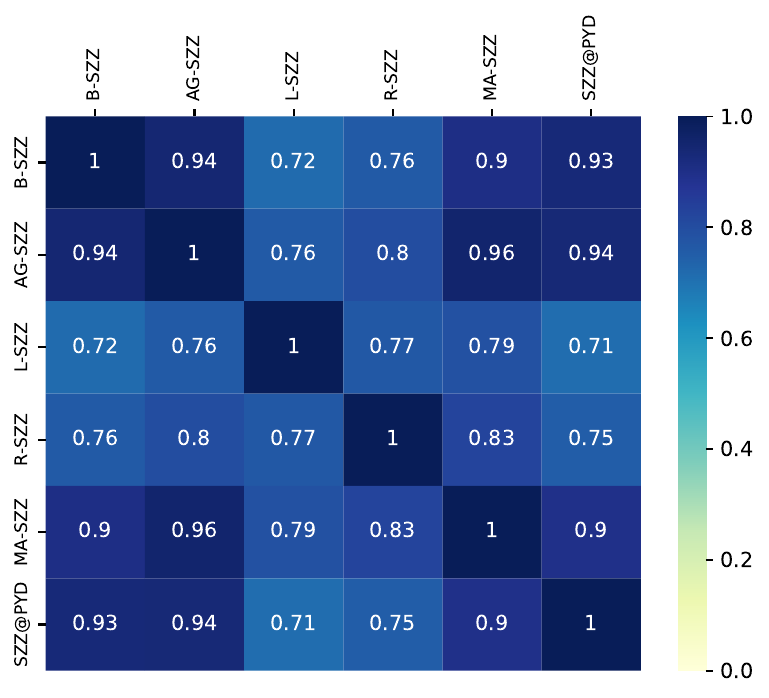}
    }
    \hfill
    \subfloat[$\text{dataset}_{\text{rosa}}$\label{fig:heatmap_pre}]{
        \includegraphics[width=0.45\textwidth]{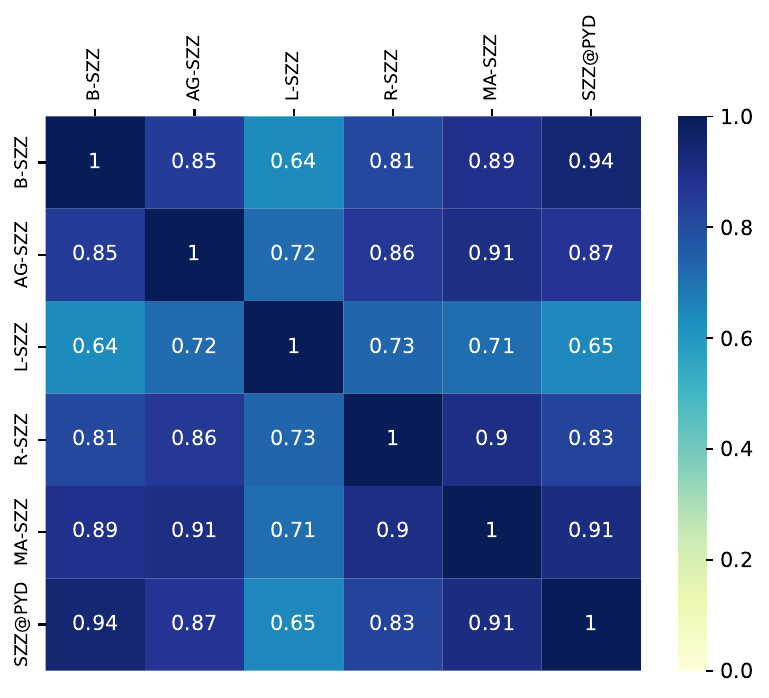}
    }
    \caption{Overlap between SZZ variants on \dataset{} dataset and $\text{dataset}_{\text{Rosa}}$}
    \label{fig:Overlap between SZZ variants}
\end{figure*}

The results depicted in Table~\ref{tab:unique} illustrate the correct identification of bug-inducing commits by technique $v_i$ that were missed by all other techniques.
Conversely, Figure~\ref{fig:heatmap} and Figure~\ref{fig:heatmap_pre} depict the $\mathit{correct}_{v_i \cap v_j}$ metric, which represents the degree of similarity between any two algorithms in their ability to accurately identify bug-inducing commits, calculated for each pair of SZZ variants on $\text{dataset}_{\text{linux}}$ and $\text{dataset}_{\text{rosa}}$ respectively.
The heatmaps exhibit symmetry, as $\mathit{correct}_{v_i \cap v_j} = \mathit{correct}_{v_j \cap v_i}$.
Figure~\ref{fig:Overlap between SZZ variants} reveals a significant overlap in the true positives identified in both datasets. 
In $\text{dataset}_{\text{rosa}}$, 13 out of 15 comparisons share more than 70\% overlap in the identified true positives, while in $\text{dataset}_{\text{linux}}$, all 15 comparisons show this level of overlap. 
\re{
For instance, in Fig~\ref{fig:heatmap}, the similarity between B-SZZ and AG-SZZ in the \dataset{} dataset can be observed at the intersection of row 2 and column 1, or equivalently at row 1 and column 2 due to symmetry, where it is 0.94.
}
The complementarity between the different SZZ variants is thus relatively low on both datasets, suggesting that there is a group of bug-fixing commits for which all variants fail to identify the correct bug-inducing commit. 

\begin{tcolorbox}
\textbf{\underline{RQ1 Findings:} Compared to Rosa's dataset, the average precision in our dataset slightly improved from 0.46 to 0.47 ($\uparrow 2.2\%$), while the average recall decreased from 0.60 to 0.52 ($\downarrow 13.8\%$). 
Similar with the findings from the previous study~\cite{rosa2021evaluating}, R-SZZ and B-SZZ remain the top performers in terms of precision and recall. 
The similarities between all the SZZ algorithms increase in the current dataset.
} 
\end{tcolorbox}

\subsection{RQ2: Ghost Commits' Impact}
\label{subsec:ghost_impact}

\begin{table}[t]
\centering
\caption{An overview of the projects and Ghost Commit' frequency. \re{RMG, AMG stand for Remove Mapping Ghost and Add Mapping Ghost.}}
\label{tab:ghost_proportion}
\begin{tabular}{lrrrr}
\hline
\textbf{Project}  &\textbf{Commits} &\textbf{RMG} & \textbf{AMG} \\ \hline
Previous Mean~\cite{rezk2021ghost}  & 12,216 & 8.05\% & 2.41\% \\
Previous Median~\cite{rezk2021ghost}  & 9,231 &  7.64\% & 2.68\% \\ 
Linux kernel (Our study)    & 1,138,001 &17.47\%  & 5.30\%\\
\hline
\end{tabular}
\end{table}

In this RQ, we aim to investigate the frequency of ghost commits in our dataset. Table~\ref{tab:ghost_proportion} shows the ghost commit frequency 
from the dataset of Rezk et al.~\cite{rezk2021ghost}, derived from 12 Apache projects, as well as our study. 
The results reveal that in the Linux kernel, 17.47\% of all bug-fixing commits are \re{Remove Mapping Ghost (i.e., bug-fixing commits involving do not have removed lines)}. 
In the previous study by Rezk et al~\cite{rezk2021ghost}, the highest percentage of \re{Remove Mapping Ghost}
observed was 11.72\%, with a median of 7.64\% across the 12 Apache projects.
Our \re{Remove Mapping Ghost} result is thus 128.66\% higher than the median from the previous study, suggesting that \re{Remove Mapping Ghost} occurrences are more common in \re{Linux kernel}, approaching one in six cases. 
This highlights the need to address this issue.

\re{Add Mapping Ghost} is also more prevalent in the Linux kernel compared to the datasets examined in prior studies. Rezk et al. reported the highest proportion of \re{Add Mapping Ghost} at 4.6\%, with a median of 2.68\%. 
However, in our \re{\dataset{}} dataset, \re{Add Mapping Ghost} constituted 5.3\% of all commits, which is 97.76\% higher than the median observed in previous research.
In Rezk et al.'s study, the ground truth of bug-inducing commits was unknown. 
Consequently, it was impossible to determine the number of bug-fixing commits for which SZZ algorithms could not identify their corresponding bug-inducing commits due to the \re{Add Mapping Ghost} problem. 
In contrast, our study has access to the ground truth, allowing us to more accurately assess the impact of \re{Add Mapping Ghost}. 
In our \re{\dataset{}} dataset, we identified 482 bug-inducing commits as \re{Add Mapping Ghost}, which appeared 585 times in 584 bug-fixing commits. 
As a result, a mere 0.77\% of bug-fixing commits are associated with an \re{Add Mapping Ghost} bug-inducing commit.
In addition, \re{Add Mapping Ghost} bug-inducing commits account for only 0.73\% of all bug-inducing commits (585 out of 79649).
However, some bug-fixing commits \re{associated with Add Mapping Ghost bug-inducing commits} experience partial success (at least one bug-inducing commit is found) or failure due to \re{Remove Mapping Ghosts}. 
In the end, for 361 bug-fixing commits (0.47\%), SZZ algorithms were unable to detect any bug-inducing commits due to \re{Add Mapping Ghosts}.

We also discovered Ghost Commits within Rosa's dataset~\cite{rosa2021evaluating}. 
Given that previous work has not studied the Ghost Commit \re{problem} in a developer-informed dataset, 
we decided to investigate the Mapping Ghosts in Rosa's dataset. 
We found that out of the 1115 bug-fixes,
only 1084 bug-fixing commits' repositories are still accessible on GitHub. 
Among these 1084 bug-fixing commits, 96 (8.86\%) are \re{Remove Mapping Ghosts}; from the 1946 bug-inducing commits, 184 (9.46\%) are \re{Add Mapping Ghosts}.

\begin{table}[t]
\centering
\caption{Results after removing the Ghost Commit. Precision, recall, and F-measure calculated for all SZZ algorithms. \re{RMG, AMG are the short for Remove Mapping Ghost and Add Mapping Ghost.}}
\label{tab:ghost_ablation}
\begin{tabular}{lccc}
\hline
\textbf{Algorithm} &\textbf{Filter RMG} &\textbf{Filter AMG} & \textbf{Filter RMG, AMG} \\ \hline
B-SZZ  & 0.70 & 0.58 & 0.71\\ 
AG-SZZ & 0.68 & 0.56 & 0.68\\ 
L-SZZ  & 0.52 & 0.44 & 0.53\\ 
R-SZZ  & 0.55 & 0.46 & 0.55\\
MA-SZZ & 0.66 & 0.55 & 0.67\\
SZZ@PYD & 0.67 & 0.56 & 0.67 \\
\hline
\end{tabular}
\end{table}

To investigate the impact of ghost commits on SZZ algorithms, we remove the \re{Remove and Add Mapping Ghost} cases from \re{\dataset{}} dataset.
\re{The experimental results for this modified dataset are presented in Table~\ref{tab:ghost_ablation}, which should be compared to the original results in Table~\ref{tab:dataset_new}.}
``Filter \re{RMG}" refers to the removal of \re{Remove Mapping Ghost} bug-fixing commits from our dataset. 
``Filter \re{AMG}" indicates the removal of \re{Add Mapping Ghost} bug-inducing commits from our dataset. 
``Filter \re{RMG, AMG}" signifies the removal of both types of commits from our dataset.
For example, when considering the B-SZZ algorithm, \re{as shown in Table~\ref{tab:ghost_ablation}}, only filtering \re{Add Mapping Ghost} reduces the recall from 0.71 to 0.58 ($\downarrow 22.4\%$), compared to filtering both \re{Remove and Add Mapping Ghost}.
While filtering only \re{Remove Mapping Ghost} leads to a decrease in recall to 0.7 ($\downarrow 1.4\%$). 
These findings suggest that \re{Remove Mapping Ghost} has a more pronounced negative impact on the recall of SZZ algorithms.
When compared to the results in Table~\ref{tab:dataset_new}, the average recall for all SZZ algorithms in the \re{\dataset{}} dataset improves by 22.2\% after the removal of \re{Remove and Add Mapping Ghost}. 
The precision of all SZZ algorithms remains unchanged due to the SZZ algorithms not producing any results for \re{Remove Mapping Ghost} cases and the relatively small proportion of \re{Add Mapping Ghost} cases.



\begin{tcolorbox}
  \textbf{\underline{RQ2 Findings:} Compared to Rezk's study, the \dataset{} dataset shows a significant increase in \re{Remove Mapping Ghosts} (from 7.64\% to 17.47\%) and \re{Add Mapping Ghosts} (from 2.68\% to 5.3\%). Notably, \re{Remove Mapping Ghosts} negatively impact the recall of the SZZ algorithms ($\downarrow 22.4\%$).}  
  \end{tcolorbox}

\subsection{RQ3: Other Failure Situations} \label{subsec:other_pattern}

Given that B-SZZ exhibits the highest recall and identifies the most unique bug-inducing commits among all the SZZ algorithms we evaluated in this paper, we further scrutinize all failure instances of B-SZZ to uncover additional failure scenarios.

Table~\ref{tab:bszz_cato} categorizes the B-SZZ algorithm results. 
``Success" indicates that B-SZZ identifies all bug-inducing commits for a bug-fixing commit; 
``Partial Success" implies at least one bug-inducing commit is identified but at least one is missed; 
``Mapping Ghost" signifies that B-SZZ can not identify the bug-inducing commits due to the Mapping Ghost Problem.
We eliminate bug-fixing commits classified as Success, Partial Success, and \re{Remove Mapping Ghost} from the \re{\dataset{}} dataset, then search for \re{Add Mapping Ghost} cases among the remaining bug-fixing commits. 
If all bug-inducing commits for a specific bug-fixing commit are \re{Add Mapping Ghost}, we consider that B-SZZ's failure is due to \re{Add Mapping Ghost}. 

Out of 584 bug-fixing commits where at least one of their bug-inducing commits is \re{Add Mapping Ghost}, 361 fail as a result of \re{Add Mapping Ghost}. 
Upon completing these filters, we obtain a dataset consisting of 17,552 bug-fixing commits where B-SZZ fails, which we refer to as "Failures without MG."

\begin{table}[t]
    \caption{Categorization of B-SZZ Algorithm Results in the Linux Kernel.}
    \centering
    \begin{tabular}{lrl}
    \hline
    Category & Number & Percentage \\ \hline
    Success & 43,846 & 57.66\% \\
    Partial Success & 1,002 & 1.32\% \\
    Mapping Ghost & 13,646 & 17.94\%  \\
    \quad \quad Remove Mapping Ghost & 13,285 & \quad 17.47\% \\
    \quad \quad Add Mapping Ghost &  361 & \quad 0.47\% \\
    Failure without MG & 17,552 & 23.08\% \\
    \quad \quad Within-File Detection & 3,965 & \quad 5.21\% \\
    \quad \quad Function-Log Detection &  4,994 &  \quad 6.57\% \\ 
    \quad \quad Line Change Detection & 3,115 & \quad 4.1\% \\
    \quad \quad Cross-File Detection & 5,478 & \quad 7.2\% \\
    \hline
    \end{tabular}
    
    \label{tab:bszz_cato}
\end{table}


With our algorithm, using  TC-SZZ, we find that out of the 17,552 bug-fixing commits in ``Failure without MG,'' 3,115 (17.7\%) of those commits contain 3,163 bug-inducing commits within the change history of some removed line. 
In contrast, {\em git log -L} only manages to track down 2,800. 
We refer to the failures that can be solved by TC-SZZ as \re{\textbf{Line Change Detection}s}.

For these 3,115 bug-fixing commits, the average length of the change history is 3.76, with the longest change history having length 35. Additionally, 1,272 bug-inducing commits occupy the second-closest position in the change history (with a total length $\geq$ 2), signifying that these bug-inducing commits can be located by using \textit{git blame} twice. 
Furthermore, 1,489 bug-inducing commits are located at the furthest point in the change history, indicating that the initial code is faulty.
The remaining 354 commits are in intermediate positions, i.e. descendant commits in Figure~\ref{fig:TC-SZZ}.

To determine if the differences in performance between TC-SZZ and the other variants of SZZ are significant, we performed a statistical analysis on the \dataset{} dataset, which comprises bug-fixing commits in the Linux kernel from 2013 to 2022. 
We divide the dataset by year to reflect the yearly increase of new bug-fixing commits.
Given the low number of bug-fixing commits in 2013 (152), we merged the data from 2013 and 2014 into a single subset for analysis.
\rem{
Initially, we used the Shapiro-Wilk test to confirm that the recall data for all algorithms follow a normal distribution. 
Subsequent t-tests were performed to compare the recall metric of TC-SZZ with other SZZ algorithms.
As shown in Table~\ref{tab:ttest_results_effect_sizes}, there are significant differences in recall between TC-SZZ and other SZZ algorithms, with all p-values significantly less than 0.05.
Furthermore, we calculated the effect size using Cohen's d~\cite{cohen2013statistical}, as reported in the fourth column of Table~\ref{tab:ttest_results_effect_sizes}. 
The effect sizes are notably large across all comparisons—with Cohen's d values exceeding 2.0—signifying substantial differences in recall between TC-SZZ and the other SZZ algorithms.}
Our analysis reveals that there is a significant difference every year between the performance of TC-SZZ and the other SZZ variants.


\begin{table}[t]
\centering
\caption{Independent t-test results comparing TC-SZZ with other SZZ algorithms in terms of recall, including effect sizes. 
}
\begin{tabular}{lccc}
\hline
Comparison & T-Statistic & P-Value & Effect Size \\ \hline
TC-SZZ vs B-SZZ & 10.3439 & $<0.0001$ & 4.876 \\
TC-SZZ vs AG-SZZ & 15.5688 & $<0.0001$ & 7.339 \\
TC-SZZ vs L-SZZ & 47.7760 & $<0.0001$ & 22.522 \\
TC-SZZ vs MA-SZZ & 18.6914 & $<0.0001$ & 8.811 \\
TC-SZZ vs PD-SZZ & 18.4269 & $<0.0001$ & 8.687 \\
TC-SZZ vs R-SZZ & 30.8874 & $<0.0001$ & 14.560 \\ \hline
\end{tabular}
\label{tab:ttest_results_effect_sizes}
\end{table}

Given that TC-SZZ can resolve 17.7\% of failure cases, we check whether we can pinpoint even more bug-inducing commits by using a TC-SZZ-like approach. 
We apply {\em git log} to all files affected by the bug-fixing commit, and then apply {\em git log -L} to all code lines within the function containing the deleted/modified lines from the bug-fixing commit.

We further define three new types of failure cases: \\
(1) \textbf{Function-Log Detection}: 
This refers to situations where the bug-inducing commit is located in the history of the function
containing deleted code, as identified using the command {\em git log -L}.
For instance, in the bug-fixing commit {\em cbcf099},\footnote{https://github.com/torvalds/linux/commit/cbcf0999ae33e7a8e\\1dba7ca935556634f679ccf} 
two lines were deleted and two lines were added. 
Despite this, neither the SZZ algorithm (B-SZZ) nor the use of {\em git blame} on \rem{all of the lines} in the function \rem{that contains} the deleted line can identify the bug-inducing commit.
However, by applying {\em git log -L} to \rem{all of the lines} in the function \rem{that contains} the deleted lines, we are able to successfully identify the bug-inducing commit.
\\
(2) \textbf{Within-File Detection}: 
This term refers to the identification of bug-inducing commits within the history of a specific file, particularly the one that was modified in a bug-fixing commit where lines of code were deleted.
For example, in commit {\em d53c51f},\footnote{https://github.com/torvalds/linux/commit/d53c51f26145657aa\\7c55fa396f93677e613548d} methods like the SZZ algorithm and the use of {\em git log -L} on the function \rem{that contains} the deleted lines \rem{can not identify} the bug-inducing commit.
\rem{Specifically, the line of code introduced by the bug-inducing commit is not the same as the one modified in the bug-fixing commit, although they are in the same file.}
In contrast, by applying git log to the entire file that encompasses these deleted lines, we can successfully uncover the bug-inducing commit.
\\
(3) \textbf{Cross-File Detection}: 
This term describes instances where the files are not shared between the bug-inducing and bug-fixing commits.
Specifically, the bug-inducing commit cannot be located within the same file as the deleted code when using the {\em git log -L} command.
\rem{As a result, since the files between bug-inducing and bug-fixing commits are not shared (as required by SZZ), it is challenging to reach such bug-inducing commits given the bug-fixing commits.}
For example, in the bug-fixing commit d15d731,\footnote{https://github.com/torvalds/linux/commit/d15d7311550983be\\97dca44ad68cbc2ca001297b} all modifications (both additions and deletions) occur in the file ``drivers/base/firmware\_loader/main.c".
However, the bug-inducing commit f531f05\footnote{https://github.com/torvalds/linux/commit/f531f05ae9437df5b\\a1ebd90017e4dd5526048e9} only modifies the file ``drivers/base/firmware\_class.c". 
This discrepancy prevents the SZZ algorithm from identifying the bug-inducing commit.
As a result, we introduce a new category of Mapping Ghosts known as "Cross-File Mapping Ghosts" to capture these scenarios where the bug-inducing changes are not found in the same file as the bug-fixing changes.

The results in Table \ref{tab:bszz_cato} reveal that out of the remaining 14,437 failure cases, 4,994 (34.6\%) are \re{Function-Log Detection}, 3,965 (27.5\%) are \re{Within-File Detection}, 
and 5,478 (37.9\%) are \re{Cross-File Detection}. 
Note that the category \re{Within-File Detection} comprises only cases that are in the file history but are not in the function history.
For the 37.9\% of failure cases that are not present in the file history, unless the bug-fixing commit explicitly mention the bug-inducing commits or specify a particular file, it becomes challenging for a researcher without domain expertise to correctly identify the bug-inducing commits. 
This task becomes even more difficult for an automated approach.


    



\begin{tcolorbox}
  \textbf{\underline{RQ3 Findings:} Our findings show that our proposed TC-SZZ can successfully address 17.7\% of failure cases. 5,478 of 14,437 (37.9\%) failure cases have bug-inducing commits that are not in the file history, making them hard to resolve.}  
  \end{tcolorbox}

\section{Classical Meets Modern: \\ Enhancing SZZ with ChatGPT}
\label{sec:discuss}


In this section, we attempt to address three types of failure situations ({\em i.e.}, \re{Function-Log Detection, Within-File Detection, and Cross-File Detection}) utilizing ChatGPT, based on the GPT-4 version, which is one of the most sophisticated Large Language Models (LLMs) currently available.

Given that ChatGPT is adept at addressing Software Engineering tasks~\cite{zhong2023can, white2023chatgpt}, we aim to investigate whether it can assist researchers or developers in more effectively identifying bug-inducing commits.
\rem{We interacted with GPT-4 through web chat using the default settings. OpenAI does not release the specific settings it employs for the web chat.}
Both the prompt and the corresponding results from ChatGPT are available in the replication package. 
We conducted the experiment from June 28th to July 4th, 2023 and from November 30th, 2023 to Jan 3rd, 2024.

Considering that GPT-4 currently has a cap of 25 messages every 3 hours and the manual process involved, we selected a sample of \re{375} failure cases\footnote{This sample size was determined to be statistically representative using a popular sample size calculator \newline (https://www.surveymonkey.com/mp/sample-size-calculator/) \newline with a confidence level of \re{95\%} and a \re{5\% margin of error}.}  
(with a \re{95\%} confidence level and \re{5\%} margin of error, \re{following previous work~\cite{hata20199, montandon2021skills, aniche2018modern, wang2021restoring}})
from the total of 14,437 identified failure cases. 
We exclude the \re{``Line Change Detection"} category because TC-SZZ was able to identify all bug-inducing commits in this group.
To simplify, we do not use TC-SZZ in this section, as our ground truth in Section~\ref{subsec:other_pattern} only uses {\em git log -L}.

In Section~\ref{subsec:other_pattern}, we classified the 14,437 failure cases into three categories: 
Function-Log Detection, Within-File Detection, and Cross-File Detection. 
This section introduces an additional category, \textbf{Function-Blame Detection}, which is a subset of {Function-Log Detection}.
The differentiation lies in the method used for identifying the bug-inducing commit. 
In the case of {Function-Log Detection}, the {\em git log -L} command is employed. 
This command helps in tracing the evolution of specific parts of a file, revealing the historical changes that led to the current state. 
For \textbf{Function-Blame Detection}, we utilize the {\em git blame} command only once. 
This command allows us to pinpoint the last change made to those lines.
\rem{There are a total of 3,200 Function-Blame Detection cases and 4,994 Function-Log Detection cases.} 
For instance, consider the commit 06bbf75.\footnote{https://github.com/torvalds/linux/commit/06bbf753476dab23e\\b262cb5fbab6d6d277a0ba3}
This commit involves a single deleted line within the function ``process\_response\_list". 
By applying {\em git blame} to all the lines of this function, we can identify the commit that initially introduced the bug. 
Although {\em git log -L} can also be used for this purpose by showing the evolution of the function over time, {\em git blame} specifically returns only the most recent changes. 


In our sample of 375 cases, we found that \rem{8.5\% (32 out of 375)} are \textbf{Function-Log Detection}s (excluding Function-Blame Detection cases), and 25.1\% (94 out of 375) are \textbf{Function-Blame Detection}s. 
Moreover, 30.9\% (116 out of 375) of the cases are \textbf{Within-File Detection}s, while 35.5\% (133 out of 375) of the cases are the \textbf{Cross-File Detection}s.

\subsection{Pipeline of ChatGPT Experiment}\label{subsec:chatgpt_pipeline}

\begin{figure}
    \centering
    \includegraphics[width=0.45\textwidth]{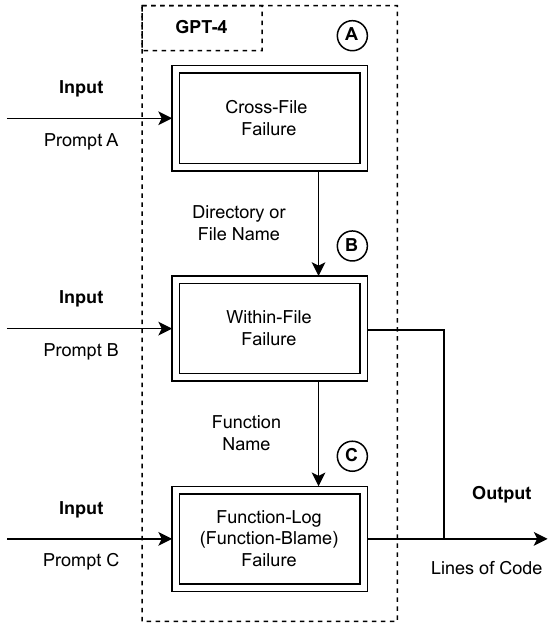}
    \caption{The Pipeline of the ChatGPT Experiment.}
    \label{fig:chatgpt_pipeline}
\end{figure}

Figure~\ref{fig:chatgpt_pipeline} shows the pipeline of the ChatGPT experiment.
The pipeline has three steps: (A) {\em Cross-File Detection}, (B) {\em Within-File Detection} (C) {\em Function-Log (Function-Blame) Detection}.
The prompts used in the different steps are shown in Table~\ref{tab:prompt}.

\begin{table}[t]
\centering
\caption{Prompt for GPT-4 in different situations.}
\begin{tabular}{|c|l|}
\hline
\textbf{Situations} & \textbf{Template} \\ \hline
\makecell[c]{\textbf{Prompt A}: \\ Cross-File \\Detection}  & \makecell[l]{〈Commit Message〉 \\ Based on the above commit message of a \\ bug-fixing commit, which file in the Linux \\ kernel could be causing this bug-fixing \\ commit?}                \\ \hline
{\makecell[c]{\textbf{Prompt B}: \\ Within-File \\ Detection }} 
& \makecell[l]{〈Commit Message〉 \\ Based on the above commit message of a \\ bug-fixing commit and function names in \\ file 〈file\_name〉, which function or functions \\ could be causing this bug-fixing commit? \\ 〈Function Name List〉 } \\ \hline
\makecell[c]{\textbf{Prompt C}: \\ Function-Log \\(Function-Blame) \\ Detection } &  \makecell[l]{〈Commit Message〉 \\ Based on the above commit message of a \\bug-fixing commit, please identify the line \\ or lines of code in the following code that \\ could be causing this bug-fixing commit: \\
〈Code of the Function〉}  \\ \hline
\end{tabular}
\label{tab:prompt}
\end{table}

For the \textbf{\re{Function-Log Detection}} 
cases, we use the Prompt C in the Table~\ref{tab:prompt} to query ChatGPT.
Specifically, for the 〈Commit Message〉 in the table, we use the commit log {\em subject} and {\em body} of the bug-fixing commit.
For 〈Code of the Function〉, we use \re{the code of all changed functions} as input for ChatGPT.
Subsequently, we execute {\em git log -L} on the line(s) of code suggested by ChatGPT to trace the bug-inducing commits.
Similarly, for the \re{\textbf{Function-Blame Detection}} 
cases, we apply the {\em git blame} command to the line(s) of code that ChatGPT indicates to identify the bug-inducing commits.
To ensure fairness in the experiment, we remove all portions of the commit message that directly reference a bug-inducing commit.

\begin{figure}
    \centering
    \includegraphics[width=0.45\textwidth]{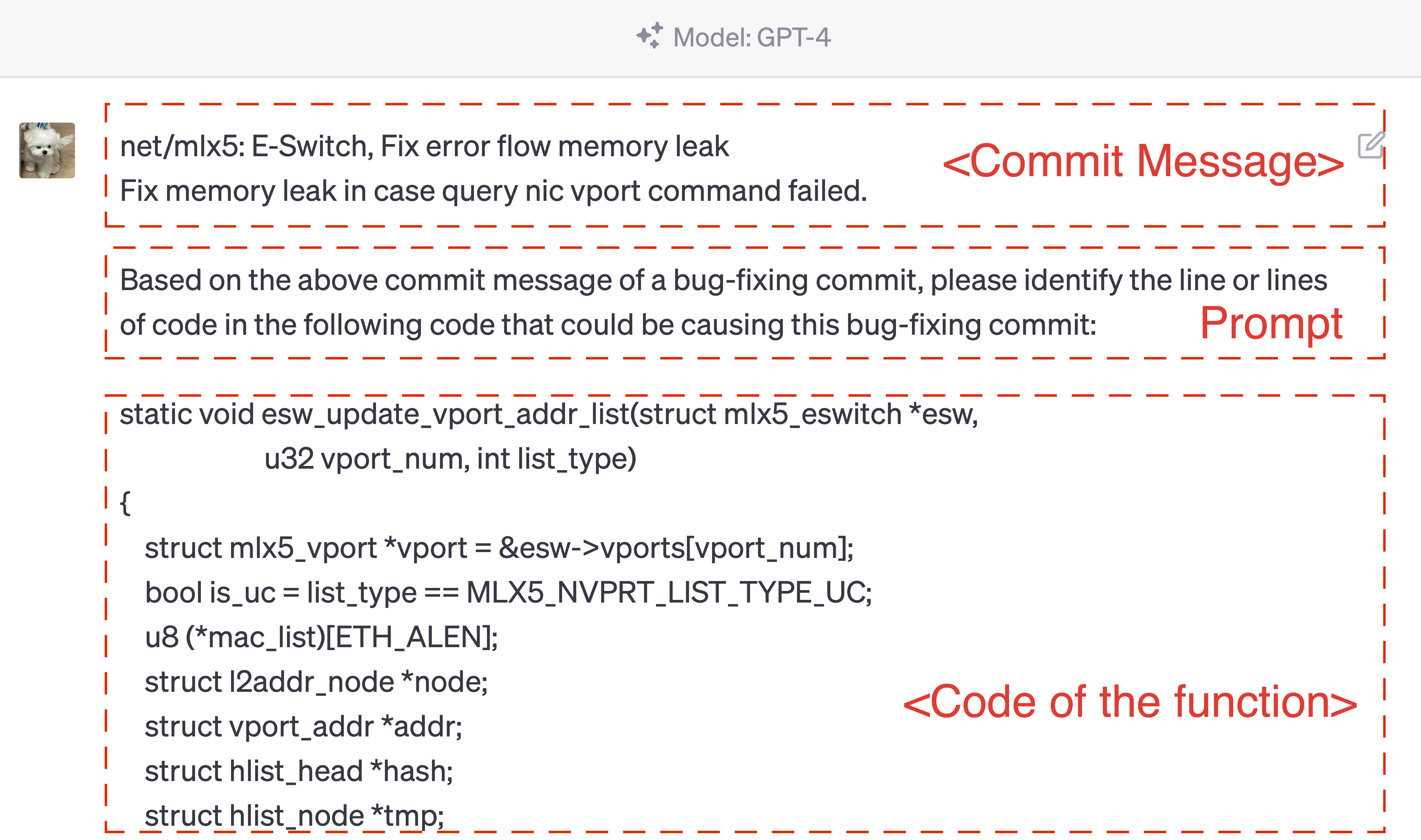}
    \caption{Input Provided to ChatGPT for Commit 761e205 in Linux kernel. In practice, we provide  the entire function code to ChatGPT, however, space constraints prevent us from including the full screenshot in the paper. The complete screenshot is available in the replication package. \\}
    \label{fig:chatgpt_input}
    
    \includegraphics[width=0.48\textwidth]{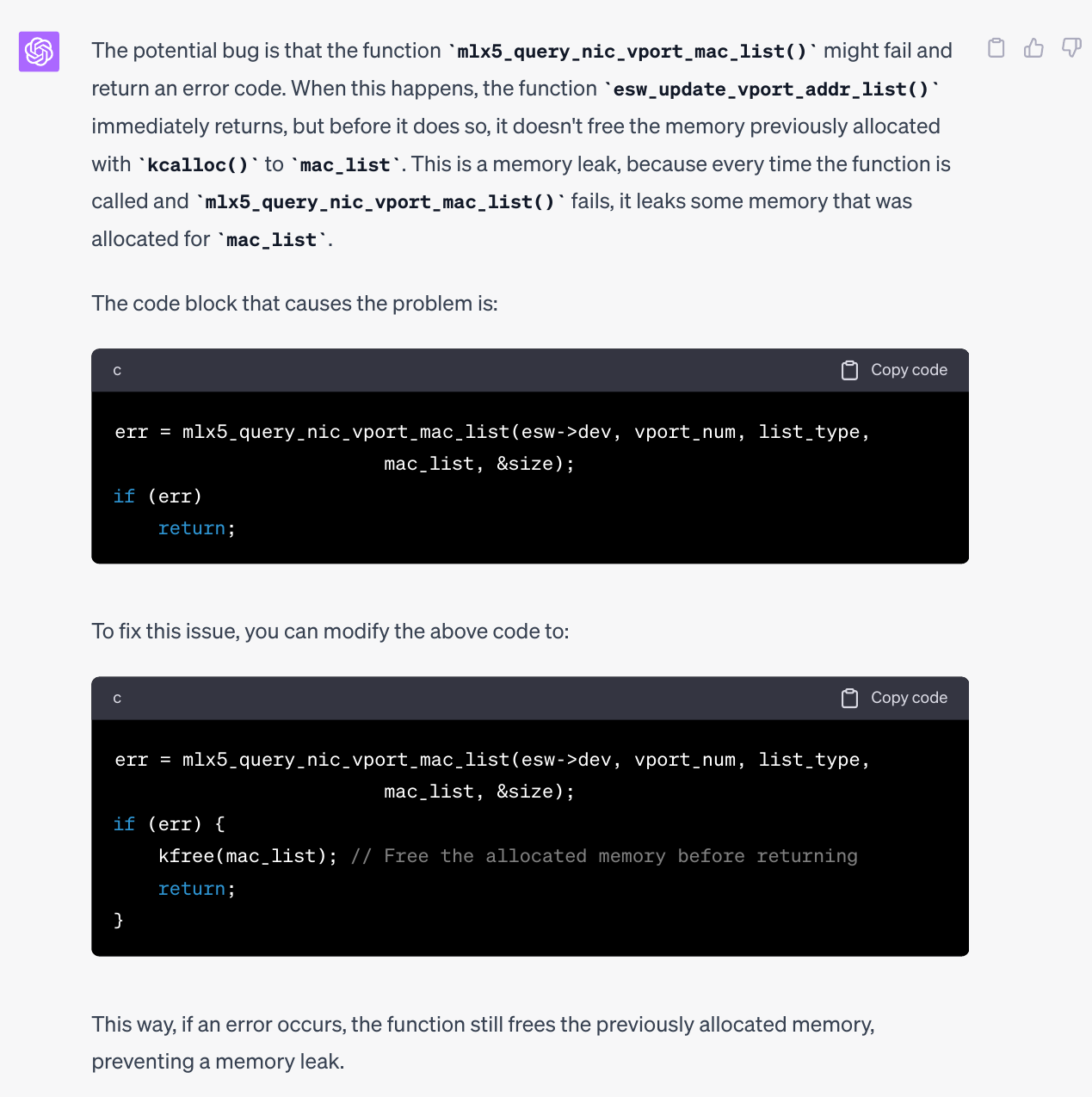}
    \caption{ChatGPT's Output for Commit 761e205 in Linux kernel. If the details in the  figure are difficult to discern, please refer to our replication package.}
    \label{fig:chatgpt_output}
\end{figure}

Figures \ref{fig:chatgpt_input} and \ref{fig:chatgpt_output} show an example where ChatGPT successfully aids in identifying a bug-inducing commit from a \re{Function-Blame Detection} case.\footnote{https://github.com/torvalds/linux/commit/761e205b559be528\\52d85e0db4a034c9f57965f9}
Initially, we query ChatGPT using the prompt depicted in Figure \ref{fig:chatgpt_input}.
Subsequently, the line(s) of code returned by ChatGPT (shown in Figure \ref{fig:chatgpt_output}) are subjected to the {\em git blame} command to get the bug-inducing commits.

For the \textbf{\re{Within-File Detection}} cases, we follow a two-step method:
(1) We first use {\em ctags}\footnote{https://ctags.io} to extract all function names in a file. 
Subsequently, we pose the Prompt B in Table~\ref{tab:prompt} to ChatGPT.
We provide ChatGPT with \re{the function names extracted using ctags}, instead of the entire code, because the number of code tokens in a file often exceeds the token limit set by ChatGPT.
(2) We then apply the same method as in the \re{Function-Log Detection} category for any correctly identified function to find the bug-inducing commits.


For the \textbf{\re{Cross-File Detection}} cases, we follow the three steps shown in Fig~\ref{fig:chatgpt_pipeline}:
(A) The initial step involves identifying the file containing the bug-inducing commits. 
We pose the Prompt A in Table~\ref{tab:prompt} to ChatGPT.
Given that the Linux kernel has 15,000 files and limited information is available in the commit message, 
ChatGPT might have difficulty identifying an exact file.  
We have observed that it could instead return a directory, despite not having been prompted to do so, containing the code affected by the bug-inducing commit. 
Hence, we evaluate ChatGPT's performance based on the correctness of the identified directory and file.
(B) If ChatGPT correctly identifies the file, we adopt the same method as the first step in the \textbf{Within-File Detection} category to determine the function. 
(3) If the correct function is determined by ChatGPT, we proceed with the same method as for the \textbf{Function-Log Detection} category to identify the bug-inducing commits.

\subsection{Results}\label{subsec:chatgpt_result}

Since the \textbf{Function-Blame Detection} category shares similar principles with the SZZ algorithm, we adopt the SZZ algorithm's evaluation methodology, as detailed in Section \ref{subsec:data_analysis}, to compute the precision, recall, and F1 for this category.
In contrast, the evaluation for the remaining three categories (i.e., \textbf{Function-Log Detection}, \textbf{Within-File Detection}, \textbf{Cross-File Detection}) diverges from the evaluation of the \textbf{Function-Blame Detection}.
Instead of adhering strictly to SZZ-based principles, for these categories, we focused on assessing the number of bug-fixing commits that could find their bug-inducing commits (i.e., Recall) with the assistance of ChatGPT.
We also used the {\em git log -L } command to identify as many bug-inducing commits as possible, aiming to evaluate the potential of a {\em git blame} based approach augmented with ChatGPT. 
For the Within-File Detection category, we evaluate three indicators (Function, Line of code, and Recall) as returned by ChatGPT.
For the Cross-File Detection cases, we evaluate five indicators: Directory, File, Function, Line of code and Recall.


The result of the ChatGPT experiment is shown in Table~\ref{tab:gpt_result}. 
For Function-Blame Detection, ChatGPT exhibits a high recall of 0.62, comparable to that of B-SZZ on the complete Linux dataset.
The precision is, however, only 0.28, because ChatGPT selects from lines of all changed functions, resulting in a substantially larger selection space, especially given that some functions span hundreds of lines.
For \textbf{Function-Log Detection}, the recall of our ChatGPT method dips to 0.31. 
Regarding \textbf{Within-File Detection}, ChatGPT successfully predicts the correct function in 32 out of 76 cases (42.1\%).
In 44 instances, ChatGPT makes an incorrect prediction, and in \rem{ten} other cases, the prediction fails due to token limitations, no functions, or ChatGPT's inability to discern the correct function.
\textbf{Cross-File Detection} presents a significant challenge as the commit messages often lack sufficient information to identify the file.
In over half the cases, ChatGPT could only return the directory of the file that might contain the bug-inducing commits. 
However, we found that ChatGPT correctly predicts the directory in 65\% of these cases. Furthermore, it successfully identifies bug-inducing commits in 11 cases where the bug-inducing commit is not in the file modified in the bug-fixing commit.
In analyzing the 11 successful cases of Cross-File Detections, we observed that the majority of the pairs of bug-fixing and bug-inducing commits are located within the same directory. 
For example, the bug-fixing commit 5b8481f and the bug-inducing commit 89a23c8 are both found within the `net/ipv6' directory. 
Nevertheless, there are four instances where the bug-fixing commit is not even in the same root directory as the bug-inducing commit.

\begin{table*}[!t]
\centering
\caption{Results of the ChatGPT Experiment. \\
// indicates that the respective indicator was not evaluated in the corresponding category.}
\label{tab:gpt_result}
\begin{tabular}{lcccccrcc}
\hline
\textbf{Failure Situations} & \textbf{Number} & \textbf{Directory} & \textbf{File}  & \textbf{Function} & \textbf{Line of Code} & \textbf{Recall} & \textbf{Precision} & \textbf{F1} \\ 
\hline
Function-Blame Detection & 94 & // & // & // & // & 0.62 & 0.28 & 0.39 \\
Function-Log Detection & 32 & // & // & // & 
\textcolor{green}{\checkmark}: 8 \textcolor{red}{\ding{55}}: 24 & 0.31 & // & // \\ 
Within-File Detection & 116 & // & // & \textcolor{green}{\checkmark}: 32 \textcolor{red}{\ding{55}}: 44 & \textcolor{green}{\checkmark}: 25 \textcolor{red}{\ding{55}}: 26 & 0.22  & // & // \\ 
Cross-File Detection  & 133 & \textcolor{green}{\checkmark}: 86 \textcolor{red}{\ding{55}}: 46 &
\textcolor{green}{\checkmark}: 43 \textcolor{red}{\ding{55}}: 46 & 
\textcolor{green}{\checkmark}: 10 \textcolor{red}{\ding{55}}: 11 &
\textcolor{green}{\checkmark}: 11 \textcolor{red}{\ding{55}}: 5 & 0.08 & // & // \\ 
\hline
\end{tabular}
\end{table*}

Through this ChatGPT experiment, we conclude that ChatGPT
exhibits potential in assisting researchers and developers in identifying bug-inducing commits. 
We acknowledge that this exploration is only preliminary, given that our experimental setup does not reflect realistic conditions, as we knew the categories beforehand. 
Nonetheless, considering that ChatGPT is a generalized model not specifically fine-tuned for this task, its performance hints at the potential benefits of similar LLMs. 
We hope that this work will catalyze further research efforts on using LLMs to overcome the limitations of the SZZ algorithm in identifying bug-inducing commits, including challenges such as ghost commits and cross-file detection.

\subsection{Evaluating ChatGPT Beyond the 2021 Data Cutoff}\label{subsec:gpt_time}

ChatGPT may have been trained on the data used in
our experiment, which could bias the results.  Our initial work with ChatGPT,
involving 165 commits, was carried out from June 28th to July 4th, 2023,
prior to OpenAI's announcement on September 27th, 2023 that it had
enhanced ChatGPT's capabilities, extending its knowledge base beyond the
previous cutoff in
September 2021.\footnote{https://twitter.com/OpenAI/status/1707077710047216095?s=20}
This raises the
opportunity to compare the success rate ChatGPT on those
commits that were released prior to September 2021 to those that were
released after that date, that were certain to be outside of the
knowledge base of ChatGPT.

\rem{In an analysis of 165 commits up to September 2023, 134 were made before September 2021, and the remaining 31 occurred after this date. 
This date serves as the data cutoff point, dividing the commits into two groups for evaluation: commits before the cutoff may have been part of the ChatGPT training data, while those subsequent to the cutoff had not.}
As shown in Table~\ref{tab:gpt_time}, for Function-Blame Detection, testing cases subsequent to the data cutoff exhibited a notable improvement with ChatGPT, achieving a 24\% increase in recall (from 0.47 to 0.71) and a 7\% increase in precision (from 0.26 to 0.33).
However, for Function-Log Detection and Within-File Detection, testing cases \rem{subsequent to} the data cutoff resulted in a 2\% decrease in performance. 
Regarding Cross-File Detection, testing cases before the data cutoff had an 8\% recall rate, whereas cases after the data cutoff failed to identify the correct bug-inducing commits in this category.
Given that testing cases after the data cutoff demonstrated a significant advantage of up to 24\% in one category, and no substantial disadvantage in the three remaining categories—with drops ranging from 2\% to 8\%—there is no conclusive evidence to suggest that testing cases potentially seen by ChatGPT would impact the experimental results.

\begin{table}[!t]
\centering
\caption{Comparative of ChatGPT on Before and After 2021 Data Cutoff.}
\label{tab:gpt_time}
\begin{tabular}{lcccc}
\hline
\textbf{Failure} & \multicolumn{2}{c}{\textbf{Before (134)}} & \multicolumn{2}{c}{\textbf{After (31)}}\\
\textbf{Situations} & \textbf{Recall} & \textbf{Pr.} & \textbf{Recall} & \textbf{Pr.}  \\ 
\hline
Function-Blame Detection & 0.47 & 0.26 & 0.71 & 0.33\\
Function-Log Detection & 0.35 & - & 0.33 & -\\ 
Within-File Detection & 0.24 & - & 0.22 & -\\ 
Cross-File Detection  & 0.08 & - & 0 & -\\ 
\hline
\end{tabular}
\end{table}

\subsection{Compare with Code-Specific Models}

To further evaluate the performance of ChatGPT, we conduct a comparative analysis with the code-specific models CodeT5+~\cite{wang2023codet5+} and CodeGen2.0~\cite{nijkamp2023codegen2}, applying a consistent experimental setting as outlined in Section~\ref{subsec:chatgpt_pipeline}.
We follow again the evaluation methodology described at the beginning of Section~\ref{subsec:chatgpt_result}.  
We use the same prompts for ChatGPT, CodeT5+, and CodeGen2.0 to ensure a fair comparison.
CodeT5+, proposed by Wang et al.~\cite{wang2023codet5+}, 
stands out as an encoder-decoder code foundation LLMs, demonstrating good performance across more than 20 code-related benchmarks involving both understanding or generative tasks, including in the zero-shot setting. 
Given that the 2B, 6B, and 16B parameter sizes
of CodeT5+ are built upon CodeGen, we also include a comparison with CodeGen2.0 in our analysis. 
However, due to constraints related to GPU memory, our assessment is confined to the codet5p-2b and codegen2-3\_7B models
for the inference stage, as these are the largest model sizes we can load.
Both models are deployed on a Docker environment with NVIDIA RTX 6000 Ada Generation (48GB of graphics memory).
\rem{
We use the same settings for the CodeT5+ 2B\footnote{https://huggingface.co/Salesforce/codet5p-2b} and CodeGen2-3.7B\footnote{https://huggingface.co/Salesforce/codegen2-3\_7B\_P} models from Hugging Face. 
The default configurations for CodeT5+ include parameters such as temperature of 1.0, model type as encoder-decoder, and 16-bit floating point tensors.
On the other hand, CodeGen2 uses similar defaults with a temperature of 1.0 but employs the GPT2Tokenizer and configures tensors to use 32-bit floating point. 
Additionally, for both models, we adjust the max\_length parameter to 2048.
}

\begin{figure}[htbp]
    \centering
    \subfloat[Same input for GPT-4, CodeT5+, and CodeGen2.0.\label{fig:input}]{
        \includegraphics[width=0.45\textwidth]{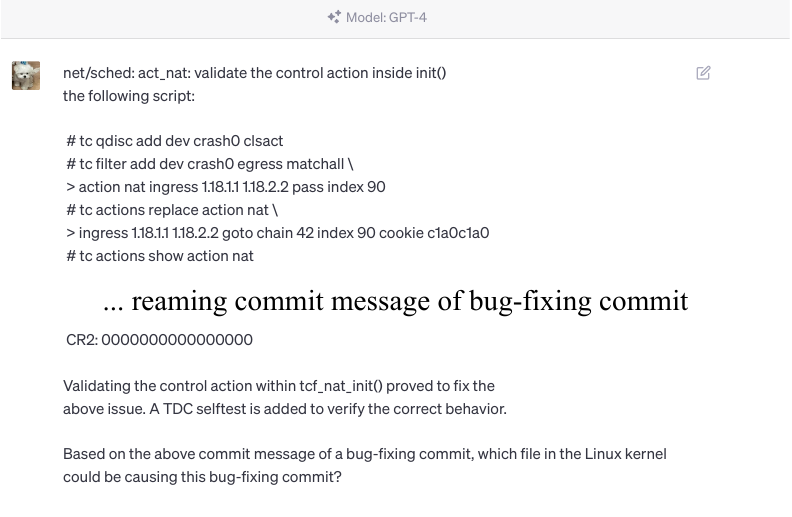}
    } \\
    \subfloat[Output of GPT-4 showing correct directory identification.\label{fig:output_gpt}]{
        \includegraphics[width=0.45\textwidth]{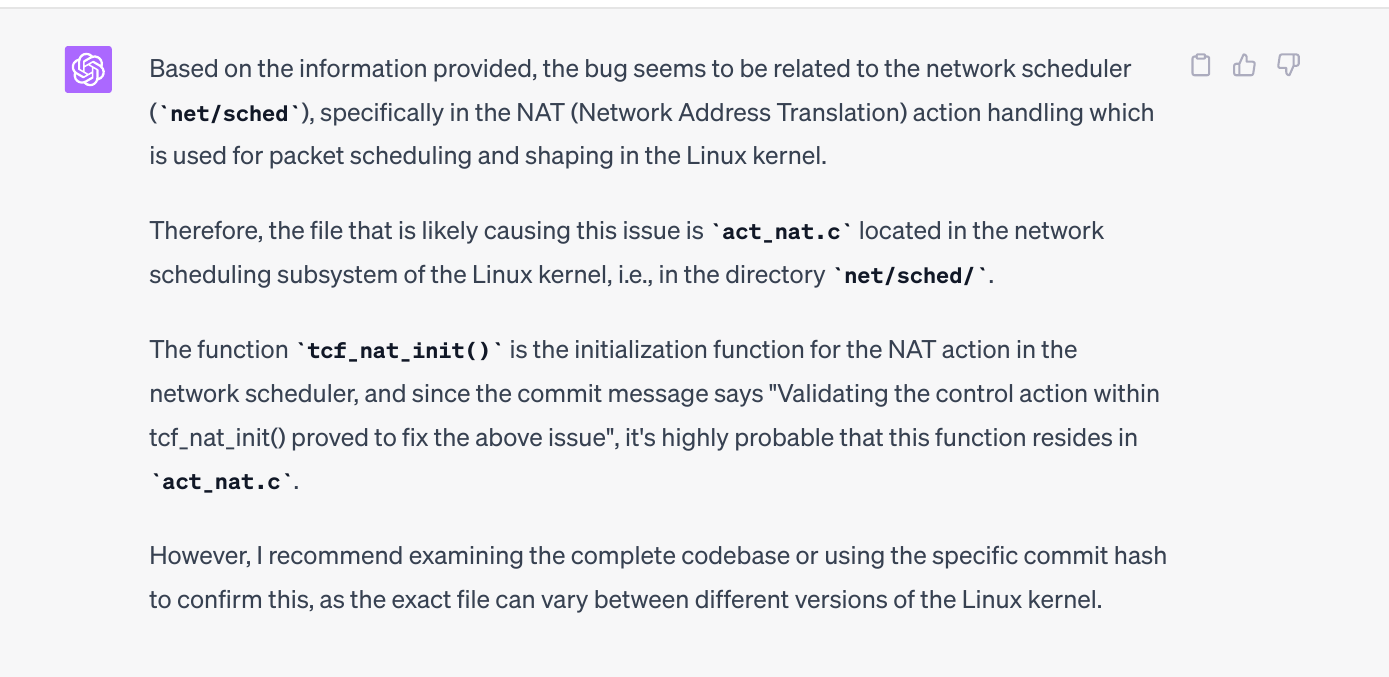}
    } \\
    \subfloat[Output of CodeT5+ displaying an error message without providing the filename.\label{fig:output_codet5p}]{
        \includegraphics[width=0.45\textwidth]{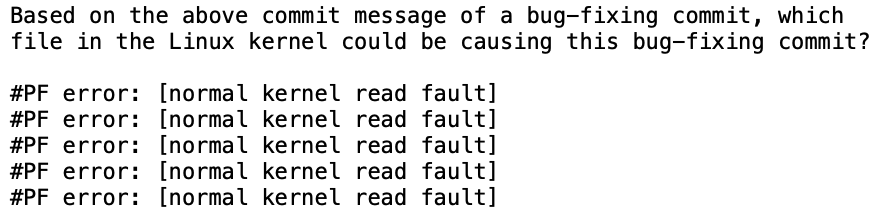}
    } \\
    \subfloat[Output of CodeGen2.0 presenting unrelated license content.\label{fig:output_codegen}]{
        \includegraphics[width=0.45\textwidth]{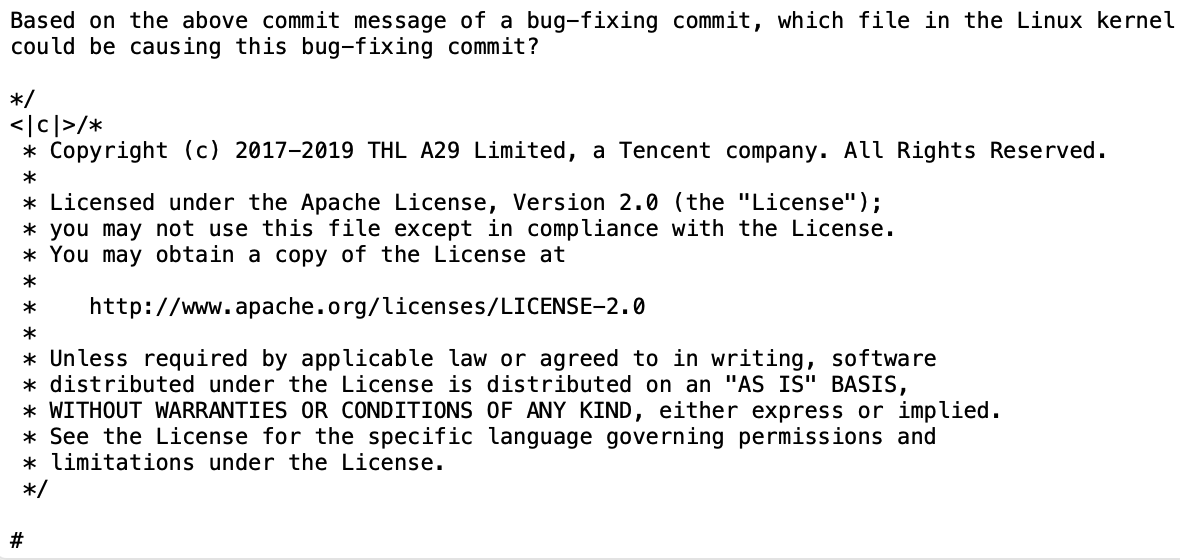}
    }\\
    \caption{Comparative Outputs for the Same Input Across Different Models: GPT-4, CodeT5+, and CodeGen2.0.}
    \label{fig:fail_case}
\end{figure}

After using the same prompt with CodeT5+ and CodeGen2.0, we found that neither model provided a reasonable answer based on the bug-fixing commit. 
For instance, when querying CodeT5+ about Cross-File Detections, where the bug-inducing commit is not in the same file as the bug-fixing commit, we sought to identify which file in the Linux kernel might be responsible. 
Unlike GPT-4, which typically suggests potential filenames or directories, CodeT5+ and CodeGen2.0 returned a nonsensical sentence. 
\rem{
For example, as shown in Fig~\ref{fig:fail_case}, all three models were prompted to identify which Linux kernel file might be the cause of the bug-fixing commit 1e45d043 (as illustrated in Fig~\ref{fig:input}).
GPT-4 correctly identified the potential directories related to the bug-inducing commit, and its response is displayed in Fig~\ref{fig:output_gpt}.
In contrast, CodeT5+ repeated responded with "\#PF error: [normal kernel read fault]" (Fig~\ref{fig:output_codet5p}), a sentence from the bug-fixing commit itself, offering no insight into the filename.
CodeGen2.0's response was also off target, returning unrelated license content, as detailed in Fig~\ref{fig:output_codegen}.}
Similarly, with Function-Blame Detections, when attempting to use prompts to have CodeT5+ or CodeGen2.0 pinpoint the problematic line, both failed to specify a line, instead generating an extensive amount of code. 

Our results demonstrate a significant limitation in the capabilities of current models, such as CodeT5+ and CodeGen, to offer precise, context-relevant assistance in identifying bug-inducing commits, particularly in complex scenarios involving cross-file analysis and specific line identification within large codebases like the Linux kernel.
It also motivates the need for developing more sophisticated LLMs, incorporating instruction tuning and domain knowledge, to improve performance in detecting bug-inducing commits, possibly even surpassing GPT-4's capabilities.

\section{Discussion}\label{sec:reason}





Although we categorized the failures of SZZ algorithms based on their location (i.e., function-log, within-file, or cross-file detection) in Section~\ref{subsec:other_pattern}, we did not explore the reasons behind these failures, which could lead to improvements in the SZZ algorithm.
In this section, we give a brief discussion based on our qualitative analysis. 

\begin{figure}[t]
    \centering
    \includegraphics[width=0.48\textwidth]{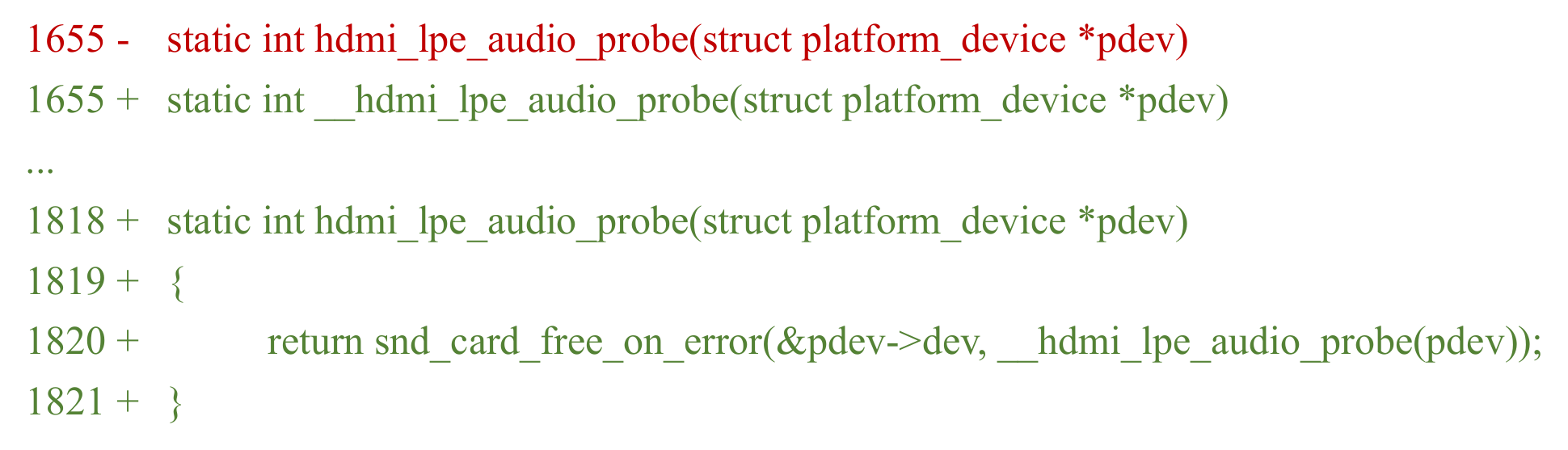}
    \caption{Example of Failure Reason: Mapping Ghost}
    \label{fig:mg3}
\end{figure}

(1) \textbf{Mapping Ghost}: This category of cases are similar to the Remove Mapping Ghost situation, where the lines removed in the bug-fixing commit do not directly assist in identifying the bug-inducing commit. 
However, they are crucial for resolving the bug.
For instance, as shown in Fig~\ref{fig:mg3}, in the bug-fixing commit 5e154df,\footnote{https://github.com/torvalds/linux/commit/5e154dfb4f9995096\\aa6d342df75040ae802c17e}
the fix creates a wrapper that causes the original function to be called via a function pointer (line 1818-1821).  
Thus, it is necessary to change the name of the original function (line 1655), and this is an integral part of the bug fix.
However, the removed line in line 1655, while crucial for the bug-fix, does not assist in identifying the bug-inducing commit. 
The core issue lies in the absence of a functionality rather than a defect in the existing functionality.

(2) \textbf{Partial Bug-fixing}: This case occurs when a bug-fixing commit rectifies a previous, incorrect bug-fix attempt, and the bug-inducing commit, as labeled by the developer, is the one that the initial, flawed bug-fix was attempting to correct. 
The SZZ algorithm identifies the recent, incorrect fix as the bug-inducing commit. 
For example, the bug-fixing commit c0702a4\footnote{https://github.com/torvalds/linux/commit/c0702a4bd41829f05\\638ec2dab70f6bb8d8010ce} addressed a bug originally introduced by commit 297ccce. 
Prior to this, commit 8299699\footnote{https://github.com/torvalds/linux/commit/829969956f97e880d\\d01086be47747226e48a3f0} attempted to fix the bug originating from commit 297ccce but failed, leading to the necessity of bug-fixing commit c0702a4.

(3) \textbf{Affected by Other Changes}: 
In some cases, code that was previously correct becomes problematic due to subsequent changes. These subsequent changes are then considered the bug-inducing commits. However, the SZZ algorithm might incorrectly identify the commit that originally wrote the code as the bug-inducing commit.
For instance, the bug-fixing commit 9b7becf1\footnote{https://github.com/torvalds/linux/commit/9b7becf103e2689d7\\f005895130ccf89a153fef1} addresses a bug induced by a previous commit.
The actual bug-inducing commit, 219fb0c, modified a function but failed to update the relevant failure message.
While the change code made in commit 9b7becf1 was appropriate at the time, it became flawed when commit 219fb0c did not make the necessary update. 
However, the SZZ algorithm incorrectly identifies the commit that initially introduced the changed code as the bug-inducing one.

\begin{figure}[t]
    \centering
    \includegraphics[width=0.48\textwidth]{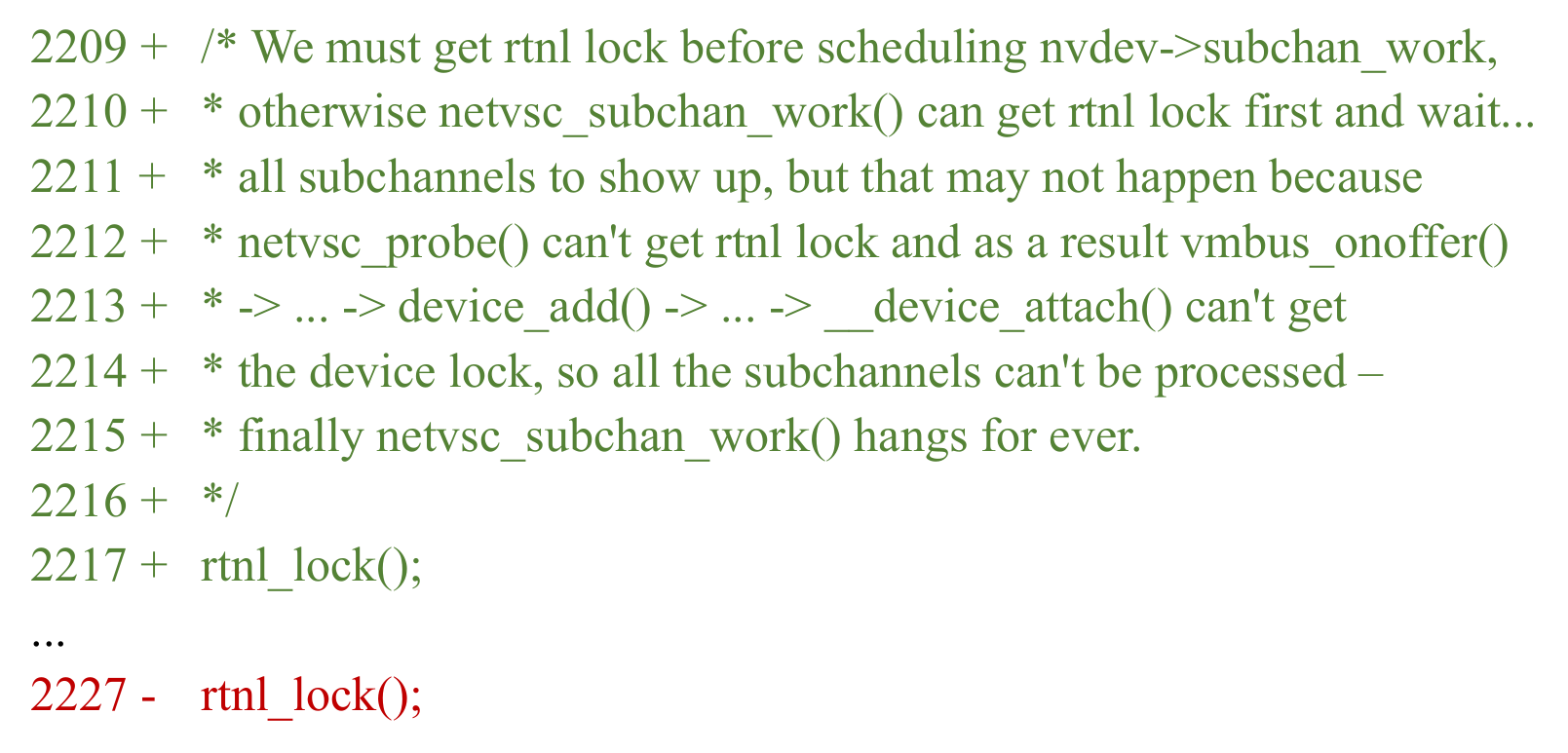}
    \caption{Example of Failure Reason: Developer Error}
    \label{fig:dev_error}
\end{figure}

(4) \textbf{Developer Error}: Even developers are not infallible and can sometimes incorrectly label commits.
For example, in the bug-fixing commit e04e7a7,\footnote{https://github.com/torvalds/linux/commit/e04e7a7bbd4bbabe\\f4e1a58367e5fc9b2edc3b10} the developer identified commit 8195b13 as the bug-inducing commit. 
However, these two commits address different aspects of deadlock issues in hv\_netvsc.
The bug-fixing commit e04e7a7 resolves a deadlock caused by a race condition between two functions and their locking mechanisms. 
In contrast, the commit labeled by the developer as bug-inducing, 8195b13, deals with a deadlock resulting channel offer messages in the same queue.
Moreover, after our review, the commit 7bf7bb3,\footnote{https://github.com/torvalds/linux/commit/7bf7bb37f16a80465\\ee3bd7c6c966f96f5a075a6} identified by the SZZ algorithm as the bug-inducing one, is considered correct.
The rationale is that in commit 7bf7bb3, rtnl\_lock() was added at an inappropriate location, leading to a deadlock. 
This can be understood from the code comments
in the bug-fixing commit, as shown in Fig~\ref{fig:dev_error}.
This revision emphasizes the role of rtnl\_lock() in preventing the deadlock and clarifies why the SZZ algorithm's identification of the bug-inducing commit is considered accurate in this context.

\section{Threats to Validity}
\label{sec:threats}

\textit{\textbf{Construct validity}}. Our dataset, collected from the Linux kernel serves as an excellent study subject.
As outlined in Section~\ref{sec:building}, we employed an automated approach to collect the data. By following the official Linux documentation for identifying bug-fixing commits, we ensured 100\% accuracy in our collected data.
A few bug-fixing commits may have been missed due to formatting errors, but this small percentage should have no impact on our results.
We used a minimum number (the commit ID must have at least 7 characters) when searching for bug-inducing commits to minimize instances where a partial commit ID matches multiple bug-inducing commits. 

\rem{As Linux Kernel does not have an issue tracker, e.g., Bugzilla or JIRA,} it is challenging to ensure that all bugs are considered.
Furthermore, the Linux documentation does not enforce the inclusion of bug-inducing commits when submitting bug-fixing commits, as the guideline suggests ``If the patch fixes a bug, cite the commit which introduced the bug if possible (and please provide both the commit ID and the title when citing commits)." 
Therefore, it is possible that some bug-fixing commits may not be linked to their corresponding bug-inducing commits, depending on the developer's discretion. 
Additionally, our dataset does not include extrinsic bugs as described by Rodriguez et al.~\cite{rodriguez2020watch}, which are bug fixes not directly traceable to changes within the repository.
As a result, our research does not claim to document every single bug-fixing commit from the past decade.
Instead, our objective is to collate as many bug-fixing commits as possible, each with a verified origin, to offer a thorough and accurate overview.

Another threat to construct validity of our dataset is that around 96\% of bug-fixing commits are linked to a single bug-inducing commit. 
This raises questions about the complexity of bugs.
The majority of 
bug-fixing commits linked to a single bug-inducing commit may be indicative of developers conforming to specific guidelines within the Linux kernel project, potentially introducing a bias into the dataset. 
Our comparison with prior research by Rosa et al.~\cite{rosa2021evaluating} revealed a similar trend: out of 1,115 bug-fixing commits, 1,113 were associated with just one bug-inducing commit.
Given these findings, it becomes challenging to definitively assert whether the reality of bug genesis is more accurately reflected in bug-fixing commits with a single or multiple bug-inducing commits. 
Therefore, we leave 
this as an open question, acknowledging the need for further discussion and exploration in this area.

In our study, we conducted the evaluation of SZZ implementations solely at the commit level, aligning with the methodology used by Rosa et al.~\cite{rosa2021evaluating}. 
It is important to note that the Linux kernel does not record pull requests (PR).
Consequently, our evaluation did not extend to the PR level.



\textit{\textbf{Internal validity}}.
There may be some pitfalls in automatically collecting bug-inducing commits. 
After selecting commits with ``Fixes:'', 2,358 (3\%) of the commits could not be matched to any bug-inducing commit.
These commit IDs are stored in a dedicated dataset.
We manually label the commit ids in this \textit{abnormal dataset} and divide them into two categories: \textit{partial commit ID} and \textit{not in the repository}.

\textit{Partial commit ID} indicates commits where the text after the keyword \textit{``Fixes:"} is not recognized by the automated filtering technique as a commit ID.
The filtering technique checks the commit ID using the regular expression \textit{[0-9a-f]\{7,40\}}, requiring 7-40 characters. 
The total number of \textit{Partial commit ID} commits is 1,021 (1.3\%). 
They include four causes: (i) the content after Fixes is a link to a bug report; (e.g., Fixes: http://bugs.gentoo.org/show\_bug.cgi?id=87182)
(ii) the content after Fixes gives some description of the bug; 
(iii) the content after Fixes has some format problems (e.g., Fixes: commit fda789fda); (iv) the commit ID after Fixes: may be too short. 

\textit{Not in the repository}
indicates that the method can recognize the commit ID in the commit message, however, the commit ID does not exist in the list of commit IDs fetched from the master branch of the Linux kernel repository.
The number of commits in this situation is 1,337 (1.7\%). 
It includes three prominent cases: (i) format problem: irregular formatting making it impossible to search for the commit ID (e.g., fda798fda:, 7598e8700e9a(drm/i915/gvt:; 
(ii) rebase problem: the commit is not in any branch of the repository, and may belong to a fork outside of the repository; (iii) branch problem: we only focus on the master branch, while some commits may occur in another branch. 

However, in total, anomalous commits only account for 3\% of the total number of commits. 
We have made these abnormal commits publicly available in a CSV file for further study.
So, we believe that the threat is minimal.

To minimize threats to internal validity from implementation errors, we use the implementation of MA-SZZ from the replication package of Rosa et al.~\cite{rosa2021evaluating} without making modifications. 
However, a deeper analysis found that the implementation of MA-SZZ by Rosa et al.~\cite{rosa2021evaluating} diverged from MA-SZZ's original implementation~\cite{da2016framework}. 
The implementation of Rosa et al. included additional modifications for detecting the movement of code within files and from other files.

Another threat to internal validity is that some bug-fixing commits are also bug-inducing commits. 
In our analysis, we found that there are 4,517 such common commits in our dataset, which represent both bug-fixing and bug-inducing actions. 
These common commits account for only 5.9\% out of the total 76,046 unique commits.
To mitigate this threat, we have listed all these commits in a file named ``common\_hashes.txt" in the replication package.
This collection of cases where developers introduce bugs while fixing bugs may provide opportunities for further study.

\textbf{\textit{External validity}}. 
Considering that our study focuses on a single software project, primarily utilizing the C programming language, our findings may not generalize to other projects or programming languages. 
However, the Linux kernel is a well-structured software repository with a substantial code size (Linux v6.0 contains 23 MLOC).
Therefore, it serves as a valuable research object for evaluating the effectiveness of SZZ algorithms.



\section{Conclusion and Future Work}\label{sec:conclusion}

SZZ is a vital algorithm for mining software repositories, and many studies rely on its results. 
Ensuring the precision and recall of SZZ in real scenarios is thus crucial. 

In our study, we collected a dataset of 76K bug-fixing commits linked to bug-inducing commits in the Linux kernel.
We evaluated six SZZ algorithms on this dataset, analyzed the ghost commits issue, and examined its impact on SZZ. 
We proposed TC-SZZ, which can find a bug-inducing commit for 17.7\% of the failure cases. 
Upon investigating the remaining failures, we found that in 37.9\% of the cases, the bug-inducing commit is not in the file history.
ChatGPT demonstrates potential in assisting the SZZ algorithm in identifying the associated bug-inducing commits in these challenging cases.

Our publicly available dataset of 76K bug-fixing commits with their bug-inducing commits provides a unique opportunity for advancing SZZ algorithm research, including training Large Language Models (LLMs) for improved precision and recall. 
Key areas for future exploration include:
(1) {\bf Enhancing SZZ Algorithm Precision:} Despite R-SZZ achieving 59\% precision, there is significant potential for improvement. 
Future studies should explore patterns beyond non-semantic source code, meta-changes, and refactoring changes to increase accuracy.
(2) {\bf Refactoring Solutions Across Programming Languages:} The extension of refactoring solutions, typically focused on Java, to other programming languages like C in our Linux dataset, remains a challenge needing attention.
(3) {\bf Addressing Undefined Failures in Recall:} Investigating the 23.08\% of previously undefined failure cases, alongside `Ghost Commit' issues, is crucial for a more comprehensive understanding and improvement of recall.
(4) {\bf Impacts on Downstream Tasks:} Utilizing our dataset to evaluate the effect of SZZ algorithms on downstream tasks, such as Just-in-Time defect prediction, offers a promising direction for future research.
These directions not only aim to refine SZZ algorithms but also seek to understand their broader implications in software engineering research.


\section{Data Availability}
The dataset used in this paper, the TC-SZZ implementation, and the results of ChatGPT experiments are publicly available at \url{https://doi.org/10.6084/m9.figshare.23889792.v2}.

\section{Acknowledgment}

This research / project is supported by the National Research Foundation, under its Investigatorship Grant (NRF-NRFI08-2022-0002). Any opinions, findings and conclusions or recommendations expressed in this material are those of the author(s) and do not reflect the views of National Research Foundation, Singapore.

\bibliographystyle{IEEEtran}
\bibliography{IEEEabrv,evaluating_szz_implementations.bib}

\begin{IEEEbiography}[{\includegraphics[width=1in,height=1.25in,clip,keepaspectratio]{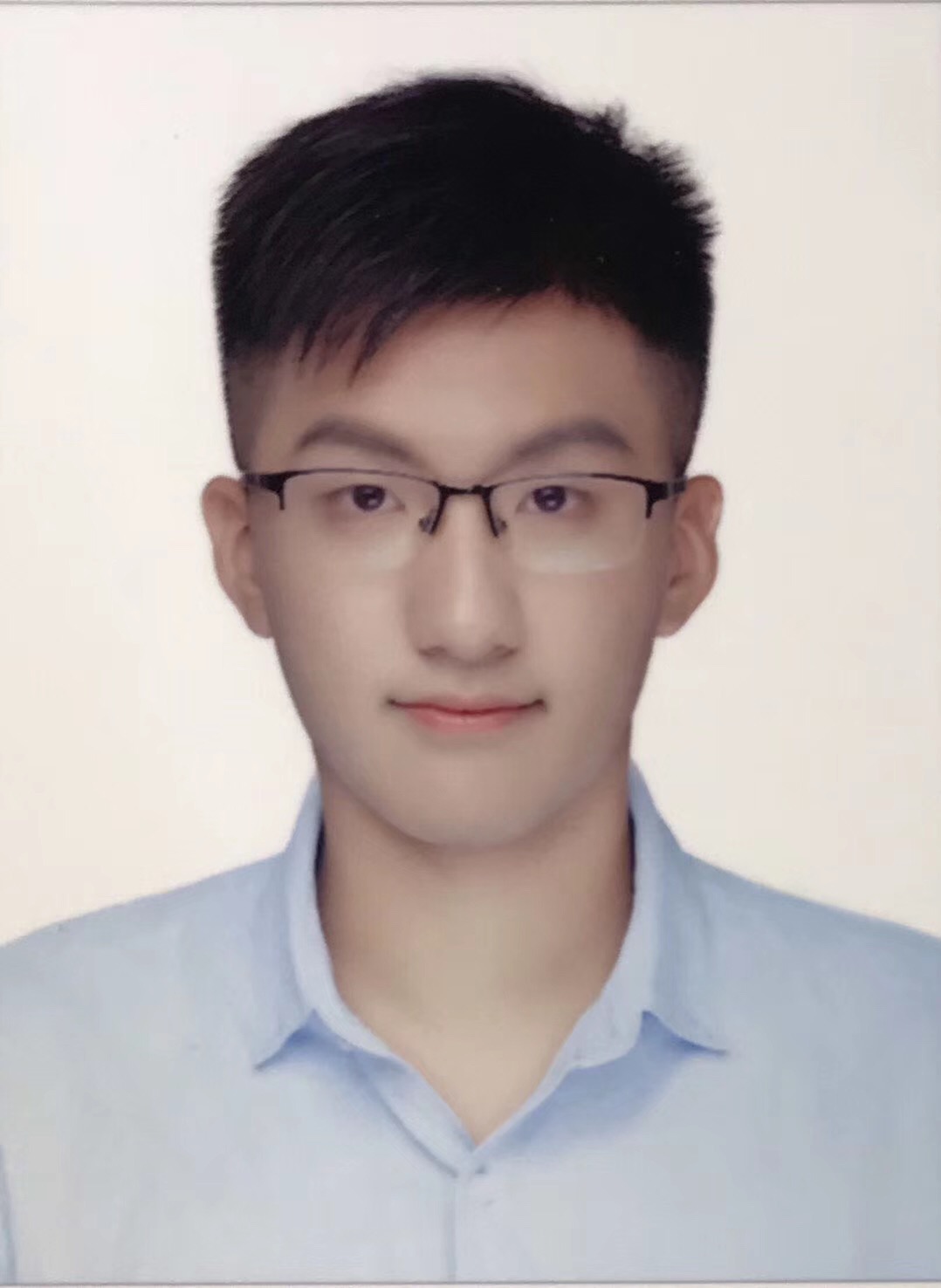}}]{Yunbo Lyu} is a Ph.D. student and a Research Engineer at the School of Computing and Information Systems, Singapore Management University under the supervision of Prof. David Lo.
Currently, his research focuses on mining software repositories and library usage.
His work has been published in high-quality software engineering conferences such as ICSE.
More information about him can be found at: https://yunbolyu.github.io.
\end{IEEEbiography}

\begin{IEEEbiography}[{\includegraphics[width=1in,height=1.25in,clip,keepaspectratio]{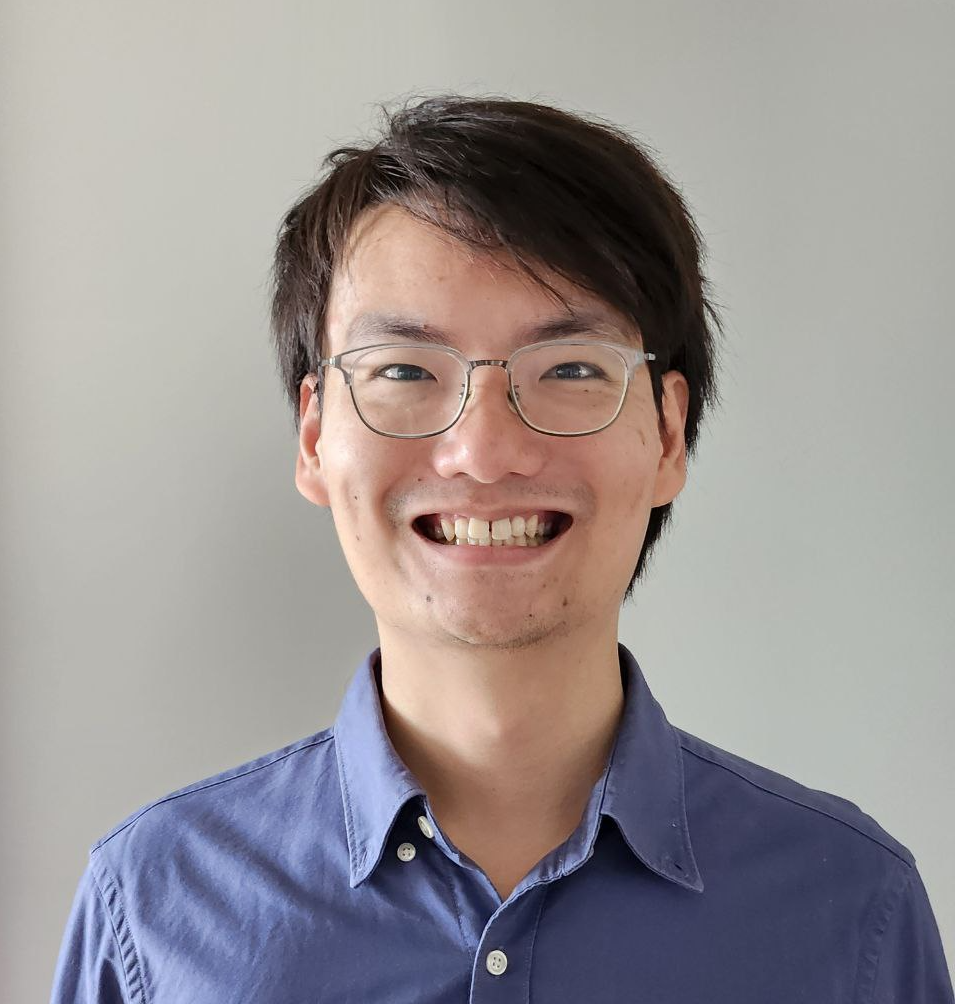}}]{Hong Jin Kang} is an incoming faculty member at the University of Sydney. He is currently a Postdoctoral Fellow at Univerity of California, Los Angeles. Previously, he completed his Ph.D. at Singapore Management University. 
His research aims to improve developer productivity by leveraging human knowledge through methods such as Active Learning. 
More information can be found at \url{https://https://kanghj.github.io/}.
\end{IEEEbiography}


\begin{IEEEbiography}[{\includegraphics[width=1in,height=1.25in,clip,keepaspectratio]{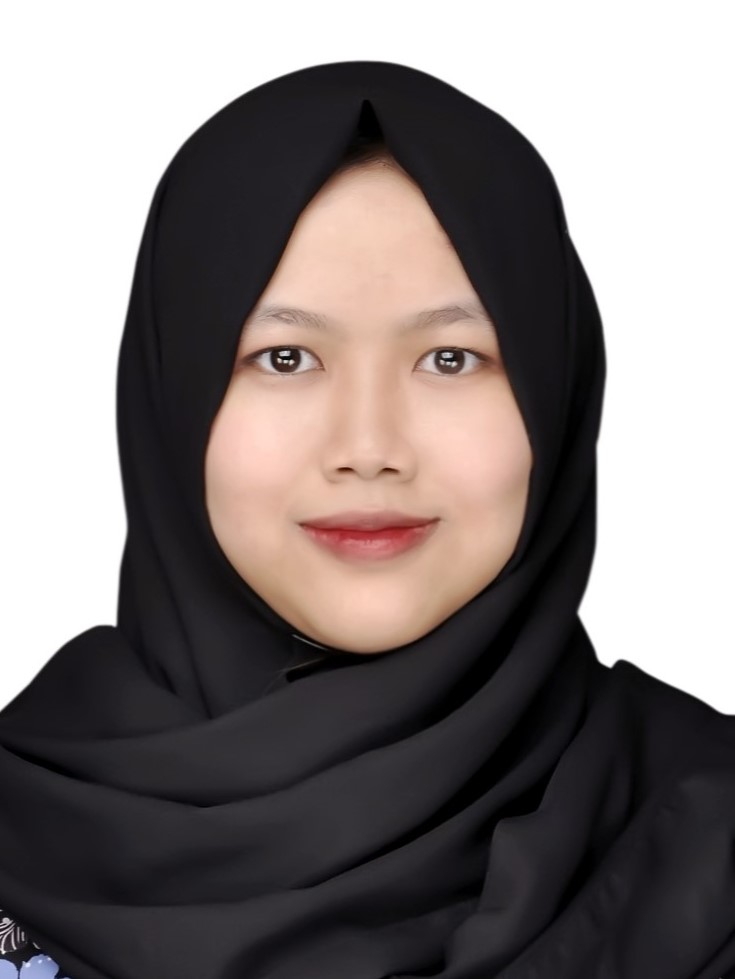}}]{Ratnadira Widyasari} is a Ph.D. candidate at the School of Computing and Information Systems, Singapore Management University under the supervision of Prof. David Lo.  Her current main research interests are in the topic of artificial intelligence and explainability for software engineering, specifically for software quality assurance. More information about her can be found at: \url{https://ratnadiraw.github.io/}.
\end{IEEEbiography}

\begin{IEEEbiography}[{\includegraphics[width=1in,height=1.25in,clip,keepaspectratio]{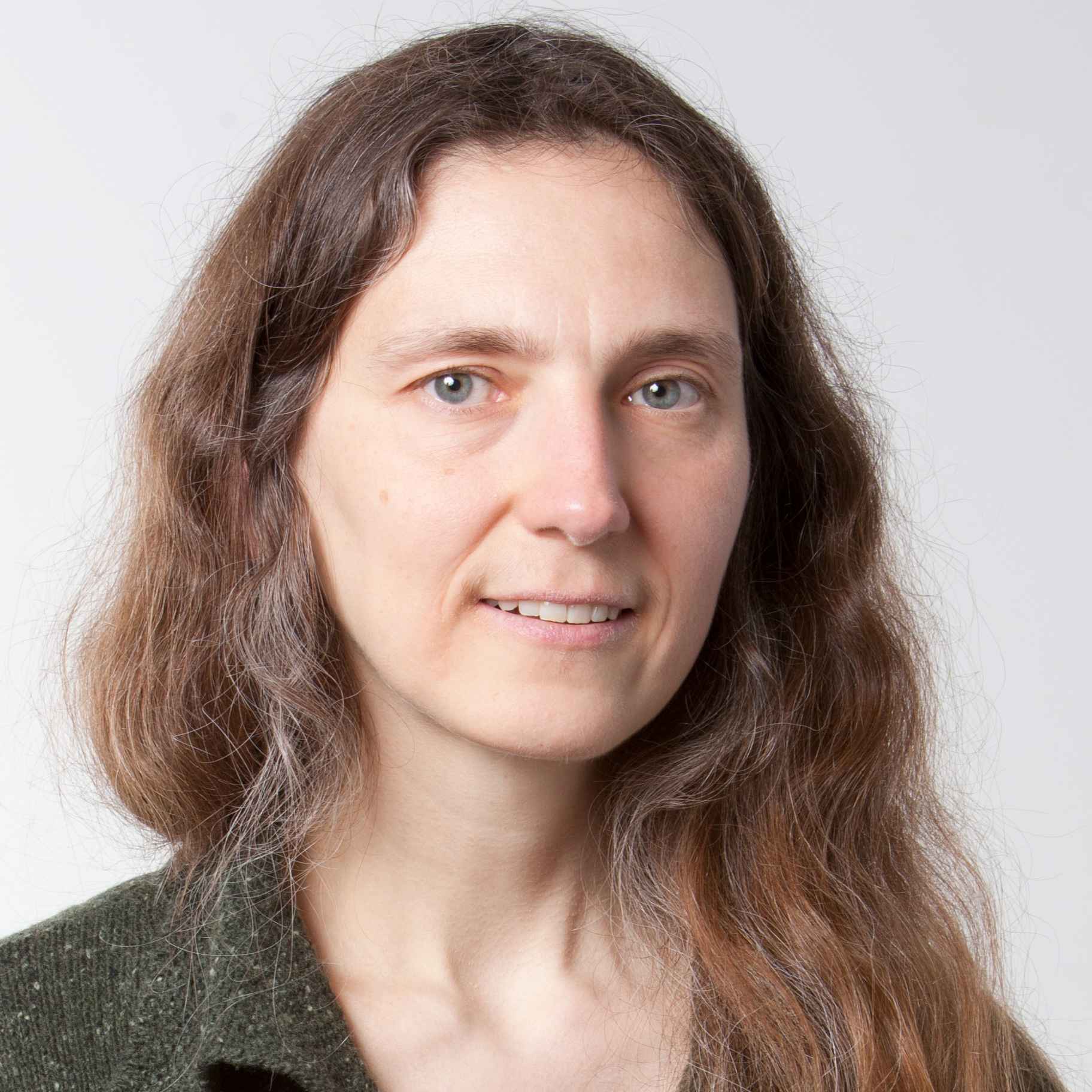}}] {Julia Lawall} is a researcher at Inria Paris, where she is the head of the Whisper team. 
She received her PhD at Indiana University in 1994.
Her current research is in the area of systems and software engineering.
Since 2004, she has been designing and maintaining the program transformation tool Coccinelle, which has been extensively used on the Linux kernel.
\end{IEEEbiography}

\begin{IEEEbiography}[{\includegraphics[width=1in,height=1.25in,clip,keepaspectratio]{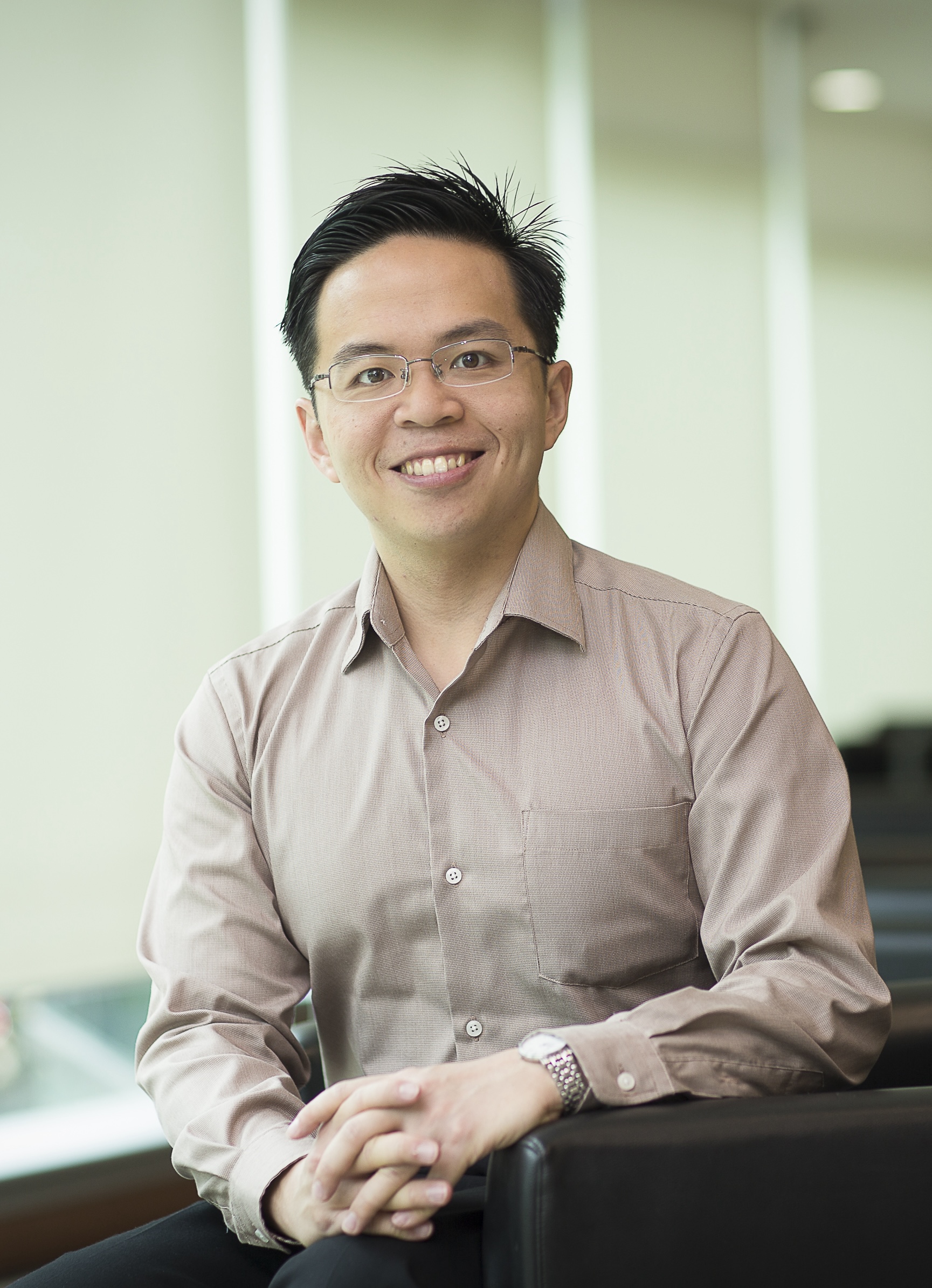}}] {David Lo} is the OUB Chair Professor of Computer Science and Director of the Center for Research in Intelligent Software Engineering (RISE) at Singapore Management University. Championing the area of AI for Software Engineering (AI4SE) since the mid-2000s, he has demonstrated how AI — encompassing data mining, machine learning, information retrieval, natural language processing, and search-based algorithms — can transform software engineering data into automation and insights. His contributions have led to over 20 awards — including two Test-of-Time awards and ten ACM SIGSOFT / IEEE TCSE Distinguished Paper awards — and gathered more than 30k citations. An ACM Fellow, IEEE Fellow, ASE Fellow, and National Research Foundation Investigator (Senior Fellow), Lo has also served as a PC Co-Chair for ASE'20, FSE'24, and ICSE'25. For more information, please visit: http://www.mysmu.edu/faculty/davidlo/
\end{IEEEbiography}

\end{document}